\documentclass[twocolumn, resetfootnote]{aastex701}

\usepackage{bm}
\usepackage{multirow}
\usepackage{float}
\usepackage{tablefootnote}
\usepackage{hyperref}

\hypersetup{
colorlinks = true,
linkcolor = green,
anchorcolor = red,
citecolor = blue,
filecolor = red,
urlcolor = red
}

\RequirePackage{color}
\definecolor{apricot}{rgb}{0.88,0.81,0.65}
\newfont{\myfont}{cmmib10}

\def\lapprox{\mathrel{\hbox{\rlap{\hbox{\lower4pt\hbox{$\sim$}}}\hbox{$<$}}}}
\def\gapprox{\mathrel{\hbox{\rlap{\hbox{\lower3pt\hbox{$\sim$}}}\hbox{$>$}}}}

\graphicspath{{./}{figures/}}

\shorttitle{Emission Properties of PSR J2129+4119 with FAST}
\shortauthors{Tedila et al.}

\begin{document}

\title{Multi-Faceted Emission Properties of PSR J2129+4119 Observed with FAST}

\correspondingauthor{H.~M. Tedila}
\email[show]{habta125@gmail.com}
\correspondingauthor{D. Li}
\email[show]{dili@tsinghua.edu.cn}
\correspondingauthor{P. Wang}
\email[show]{wangpei@nao.cas.cn}

\author[orcid=0000-0002-5815-6548,gname=Habtamu,sname=Tedila]{H.~M. Tedila}
\affiliation{National Astronomical Observatories, Chinese Academy of Sciences, A20 Datun Road, Chaoyang District, Beijing 100101, People's Republic of China}
\affiliation{Arba Minch University, Arba Minch 21, Ethiopia}
\email{habta125@gmail.com}

\author[orcid=0000-0003-3010-7661,gname=Di,sname=Li]{D. Li}
\affiliation{New Cornerstone Science Laboratory, Department of Astronomy, Tsinghua University, Beijing, People's Republic of China}
\affiliation{National Astronomical Observatories, Chinese Academy of Sciences, A20 Datun Road, Chaoyang District, Beijing 100101, People's Republic of China}
\affiliation{State Key Laboratory of Radio Astronomy and Technology, Beijing 100101, People's Republic of China}
\email{dili@tsinghua.edu.cn}

\author[orcid=0000-0002-3386-7159,gname=Pei,sname=Wang]{P. Wang}
\affiliation{National Astronomical Observatories, Chinese Academy of Sciences, A20 Datun Road, Chaoyang District, Beijing 100101, People's Republic of China}
\affiliation{Institute for Frontiers in Astronomy and Astrophysics, Beijing Normal University, Beijing 102206, People's Republic of China}
\affiliation{State Key Laboratory of Radio Astronomy and Technology, Beijing 100101, People's Republic of China}
\email{wangpei@nao.cas.cn}

\author[gname=Rai,sname=Yuen]{R. Yuen}
\affiliation{Xinjiang Astronomical Observatory, Chinese Academy of Sciences, 150 Science 1-Street, Urumqi, Xinjiang, 830011, People's Republic of China}
\affiliation{Xinjiang Key Laboratory of Radio Astrophysics, 150 Science1-Street, Urumqi, Xinjiang, 830011, People's Republic of China}
\email{ryuen@xao.ac.cn}

\author[orcid=0000-0002-1381-7859,gname=Zi-Wei,sname=Wu]{Z.~W. Wu}
\affiliation{National Astronomical Observatories, Chinese Academy of Sciences, A20 Datun Road, Chaoyang District, Beijing 100101, People's Republic of China}
\email{wuzw@bao.ac.cn}

\author[gname=Shu-Jin,sname=Dang]{S.~J. Dang}
\affiliation{School of Physics and Electronic Science, Guizhou Normal University, Guiyang, 550001, People’s Republic of China}
\email{dangsj@gznu.edu.cn}

\author[orcid=0000-0002-5381-6498,gname=Jian-Ping,sname=Yuan]{J.~P. Yuan}
\affiliation{Xinjiang Astronomical Observatory, Chinese Academy of Sciences, 150 Science 1-Street, Urumqi, Xinjiang, 830011, People's Republic of China}
\affiliation{Xinjiang Key Laboratory of Radio Astrophysics, 150 Science1-Street, Urumqi, Xinjiang, 830011, People's Republic of China}
\email{yuanjp@xao.ac.cn}

\author[orcid=0000-0002-9786-8548,gname=Na,sname=Wang]{N. Wang}
\affiliation{Xinjiang Astronomical Observatory, Chinese Academy of Sciences, 150 Science 1-Street, Urumqi, Xinjiang, 830011, People's Republic of China}
\affiliation{Xinjiang Key Laboratory of Radio Astrophysics, 150 Science1-Street, Urumqi, Xinjiang, 830011, People's Republic of China}
\email{na.wang@xao.ac.cn}

\author[gname=Marilyn,sname=Cruces]{M. Cruces}
\affiliation{Centre of Astro-Engineering, Pontificia Universidad Cat\'olica de Chile, Av. Vicu\~na Mackenna 4860, Santiago, Chile}
\affiliation{Department of Electrical Engineering, Pontificia Universidad Cat\'olica de Chile, Av. Vicu\~na Mackenna 4860, Santiago, Chile}
\email{mscruces@uc.cl}

\author[gname=Shuo-Jun,sname=Zhang]{J. S. Zhang}
\affiliation{National Astronomical Observatories, Chinese Academy of Sciences, A20 Datun Road, Chaoyang District, Beijing 100101, People's Republic of China}
 \affiliation{University of Chinese Academy of Sciences, 19A Yuquan Road, 100049 Beijing, People's Republic of China}
\email{zhangjs@bao.ac.cn} 

\author[gname=Juntao,sname=Bai]{J. Bai}
\affiliation{Institute for Gravitational Wave Astronomy, Henan Academy of Sciences, Zhengzhou 450046, Henan, People’s Republic of China}
\email{jtbai5537@gmail.com}

\author[gname=De,sname=Zhao]{D. Zhao}
\affiliation{Xinjiang Astronomical Observatory, Chinese Academy of Sciences, 150 Science 1-Street, Urumqi, Xinjiang, 830011, People's Republic of China}
\email{zhaode@xao.ac.cn}

\collaboration{all}{FAST Collaboration}

\begin{abstract}
We present a detailed single-pulse study of the long-period pulsar PSR~J2129+4119 using high-sensitivity FAST observations. Despite locating well below the traditional death line, the pulsar exhibits sustained and multi-modal emission behavior, including nulls, weak pulses, regular emission, and occasional bright pulses. The nulling fraction is measured to be $8.13\% \pm 0.51\%$, with null durations typically under four pulse periods. Fluctuation spectral analysis reveals both phase-modulated subpulse drifting and intermittent beat-like modulation. At the same time, polarization profiles show high linear polarization and stable polarization position angle (PPA) swings consistent with a near-tangential sightline geometry. Quasi-periodic microstructures are detected in 11.54\% of regular pulses, with a mean periodicity and width of 4.57 ms and 4.30 ms, respectively. A well-defined scintillation arc in the secondary spectrum confirms the presence of a localized scattering screen. These results indicate that PSR~J2129+4119 remains magnetospherically active and coherently emitting despite its low energy loss rate, offering key insights into pulsar emission physics near the death line.\\
\end{abstract}


\setcounter{footnote}{0}

\section{INTRODUCTION}\label{sec:intro}

Since the first discovery in 1967 \citep{Hewish1968}, over 3,700 pulsars have been identified\footnote{\url{https://www.atnf.csiro.au/research/pulsar/psrcat/}}. Each pulsar exhibits a stable (with few exceptions) and unique average pulse profile formed by integrating thousands of individual pulses at a given frequency \citep{Helfand1975, Lyne1988}. However, the emission mechanism remains poorly understood, and single-pulse studies are essential for probing the underlying physics. These observations reveal a rich diversity of variability phenomena, including subpulse drifting \citep{Drake68, Yan2023, Janagal2023}, nulling \citep{Backer70, Wang07, Rahaman2021, HMT2022}, mode changing \citep{Wang07, Rahaman2021, Shang2024}, and long-term amplitude modulation \citep{Brook2019, Song2023, Hsu2025}. Such behaviors are particularly common in long-period pulsars \citep[e.g., PSRs J1945+1211, J2323+1214, and J1900$-$0134;][]{HMT2025}. These phenomena offer valuable insight into the plasma conditions and electrodynamic processes within the pulsar magnetosphere \citep[e.g.,][]{Melrose14, Mitra2024}.

One of the most extensively studied pulsar phenomena is pulsar nulling, which is characterized by the sudden cessation of pulsed emission lasting from a single pulse to thousands of pulse periods \citep{Wang07}. First reported by \citet{Backer70}, nulling has since been identified in more than 200 pulsars \citep{Gajjar14, Sheikh2021, HMT2025}, with nulling fractions (NF) ranging from less than 1\% to more than 90\% \citep{Ritchings76, Wang07}. Nulls are often interpreted as interruptions in pair production or changes in magnetospheric conductivity that suppress coherent emission \citep{Timokhin10, Timokhin2019, Basu17}. In some pulsars, such as PSR B1931+24, nulling is accompanied by changes in spin-down rate, suggesting a restructuring of global magnetospheric currents \citep{Kramer06}. Closely related to nulling is the phenomenon of mode changing, where the pulsar abruptly switches between two or more stable emission states, each with distinct pulse profile shape, intensity, or subpulse behavior \citep{Wang07, Shang2024}. While nulling involves a complete cessation of emission, mode changing retains radio output but with altered properties. Despite these differences, both phenomena are likely to reflect transitions between distinct magnetospheric states involving large-scale changes in current flow and plasma distribution \citep{Mitra2024}. 

Another prominent form of single-pulse modulation is subpulse drifting, in which subpulses systematically shift in phase across successive rotations \citep{Drake68}. This behavior is well explained by the carousel model, wherein plasma subbeams circulate around the magnetic axis due to $\mathbf{E} \times \mathbf{B}$ drift in the acceleration region \citep{Ruderman1975, Deshpande01}. This phenomenon is described by two key periodicities: $P_2$, the angular separation between adjacent subpulses, and $P_3$, the interval over which subpulse patterns repeat at a fixed pulse phase. Spectral analyses of these parameters provide valuable constraints on the geometry and electrodynamic processes in the inner magnetosphere \citep{Weltevrede06, Basu16, Basu19, Song2023}. Some pulsars exhibit multiple drift modes, abrupt changes in drift rate, or coupling between drifting and nulling \citep{Rankin86, vLetal03, McSweeney2019}, suggesting a unified framework that links various types of emission variability. For instance, PSR~J0344$-$0901 exhibits four distinct emission modes, each characterized by unique subpulse movement \citep{HMT2024}; PSR~J1921+1948 shows three drifting modes along with substantial nulling and mode-changing behavior \citep{Shang2024}; and PSR~B2319+60 presents two distinct drift modes in addition to a phase-stationary, non-drifting emission mode, each associated with a different pulse profile \citep{Chen2022}.

In addition to nulling and drifting, some pulsars show long-term amplitude modulations or beat-like variability, where the pulse energy fluctuates quasi-periodically over tens to hundreds of rotations \citep{Janagal2023, Hsu2025}. These modulations may result from interference between closely spaced drift frequencies, periodic carousel rotation, or long-timescale instabilities in the magnetosphere \citep{Weltevrede06, Rankin08, Szary2024}. For example, PSR~J1514$-$4834 displays drifting subpulses alongside rapid periodic amplitude modulation \citep{Hsu2025}, similar to PSR~B0943+10, which shows sideband features interpreted as a beat pattern from a slowly circulating carousel of subbeams \citep{Deshpande1999, Gil2003}. Modulation effects linked to carousel dynamics have also been observed in PSR~B0834+06  \citep{Asgekar2005}, PSR~B1857$-$26 \citep{Mitra2008}, and PSR~J2022+5154 \citep{Chen2024}. Although relatively rare, these beat-like modulations offer valuable insights into subbeam structure and the evolving physical conditions within pulsar magnetospheres \citep{Dyks2021}.

The coherent radio emission observed from pulsars is generally attributed to curvature radiation from bunches of charged particles accelerated along open magnetic field lines \citep{Ruderman1975, Rahaman20}. Primary particles initiate pair cascades, generating dense secondary plasma that sustains the radio emission \citep{Timokhin2019, Philippov2020}. Electric fields parallel ($E_\parallel$) and perpendicular ($E_\perp$) to the magnetic field lines further modulate the emitted signal, resulting in high linear polarization and frequency-dependent structure \citep{Gil2002, Wang2015}.

In this paper, we present the first high-sensitivity single-pulse and timing study of PSR~J2129+4119 using observations from the Five-hundred-meter Aperture Spherical Radio Telescope (FAST). The pulsar was discovered on 2017 October 10 during pilot scans of the Commensal Radio Astronomy FAST Survey (CRAFTS; \citealt{Li18})\footnote{\url{https://crafts.bao.ac.cn/pulsar/}}, using the ultrawide-bandwidth (UWB) receiver and the Parkes Digital Filter Bank backend (PDFB4). It was later included in a timing campaign with the 100-m Effelsberg telescope at 1.36\,GHz \citep{Cruces2021}, but no single-pulse analysis was conducted. Our FAST observations reveal a rich variety of emission behaviors, including nulling, subpulse drifting, and beat-like amplitude modulations. With a pulse period of 1.69\,s, a dispersion measure (DM) of 31\,cm$^{-3}$\,pc, and a characteristic age of 342.8\,Myr \citep{Cruces2021}, PSR~J2129+4119 is an old pulsar exhibiting multiple distinct emission modes. It is located below the traditional death line in the \( P \)–\( \dot{P} \) diagram, making it a valuable object for investigating emission physics near the death boundary. 
The paper is organized as follows. Our observation setup and data processing methods are described in Section~\ref{sec:observe}. The analysis of pulse sequences and classification of emission modes is presented in Section~\ref{sec:analysis_result}. Frequency-dependent pulse profile variations and subpulse drifting behavior are discussed in Sections~\ref{sec:freq-dependence} and~\ref{sect:subpulse_drifting}, respectively. Finally, Sections~\ref{sec:discus} and~\ref{sec:Summary} present the discussion of our findings and a summary of the main emission features of the pulsar.

\section{OBSERVATION AND DATA PROCESSING}\label{sec:observe}

FAST is the largest and most sensitive single-dish radio telescope in the world. It is a national mega-science project located in Guizhou Province, China, at geographic coordinates $25^\circ.7$\,N, $106^\circ.9$\,E \citep{Nan11,Li18}. FAST has a 500-m diameter reflector, with a 300-m illuminated aperture actively used during observations. The main structure of the telescope was completed on 2016 September 15, and it entered the commissioning phase, which lasted from September 2016 to May 2018 \citep{Jiang19,Jiang20}. During this phase, the telescope employed an ultra-wideband (UWB) receiver covering 270--1620\,MHz \citep{Cameron20}. Since May 2018, it has operated with a 19-beam L-band receiver spanning 1.0--1.5\,GHz.

Observations of PSR~J2129+4119 were carried out using the 19-beam receiver and the Reconfigurable Open Architecture Computing Hardware Version-2 (ROACH2) signal processor \citep{Jiang19}, in two modes: single-pulse observation and timing observation. As the timing behavior of this pulsar has previously been studied by \citet{Cruces2021}, we employed the pulsar ephemeris provided by the Australia Telescope National Facility Pulsar Catalogue \citep[ATNF Pulsar Catalogue;][]{Manchester05}.

The single-pulse observation was performed on 2024 September 29 for one hour, at a central frequency of 1.25\,GHz with a 400\,MHz bandwidth. The data were recorded in 8-bit search-mode PSRFITS format \citep{Hotan04} with 1024 frequency channels, and time and frequency resolutions of 49.152\,$\mu$s and 0.488\,MHz, respectively. The raw data were processed with the \texttt{DSPSR} package to generate single-pulse sequences \citep{Straten11}. As pulsar observations are often affected by narrowband non-pulsar radio-frequency interference (RFI), both automatic and manual RFI mitigation were carried out using the Pulsar Archive Zapper (\texttt{PAZI}) and \texttt{PAZ} routines in \texttt{PSRCHIVE}\footnote{https://psrchive.sourceforge.net} \citep{Hotan04}. The cleaned and folded single-pulse data were then analyzed using \texttt{PSRSALSA}\footnote{http:// psrchive.sourceforge.net/ manuals/ psrspa/} \citep{Weltevrede16} to investigate subpulse drifting, nulling, and mode-changing behavior.

A timing observation campaign was conducted regularly between 2023 July 07 and 2024 July 03, with each session lasting approximately five minutes. To maintain consistency across epochs, all data were processed using the \texttt{pam} utility with a common phase reference (\(\phi_0\)), ensuring uniform on-pulse phase alignment across all observations. During each observation, a stable noise signal was injected using a calibration diode, and the resulting calibration file was folded at a reference period of 0.1006632960\,s.
Polarization calibration was then carried out using the Pulsar Archive Calibration (\texttt{PAC}) tool in PSRCHIVE with a calibrator database based on known sources. Calibrators were matched by time and sky position, while matching by instrument and frequency was disabled. Frontend corrections were omitted, and the sign of Stokes~$Q$ was flipped using the \texttt{-nq} option to ensure correct polarization orientation. This calibration procedure follows the approach of \citet{Yan11}. The rotation measure (RM) was subsequently determined using \texttt{rmfit}, which optimizes the linear polarization fraction over a trial range of $[-10000, 10000]$\,rad\,m$^{-2}$ in 100\,rad\,m$^{-2}$ steps. We obtained an RM of $-47.02 \pm 4.65$\,rad\,m$^{-2}$ using \texttt{rmfit}, which includes ionospheric contributions and was used to correct the polarization profiles. This value differs from the RM of $-30 \pm 0.9$\,rad\,m$^{-2}$ reported by \citet{Cruces2021}, which was derived using the RM synthesis method described in \citet{Brentjens05}. While earlier trials using \texttt{rmfit} generally suggested values with $|\mathrm{RM}| < 300$\,rad\,m$^{-2}$, our calibration and fitting yielded a more constrained estimate. These corrections were essential for reliable RVM fitting and mode-dependent polarization analysis. 

Since our observations were not flux-calibrated, we estimated the mean flux density using the radiometer equation \citep{LK04} as follows:
\begin{equation}
	S_{\rm mean} = \frac{\beta\, (S/N)\, T_{\rm sys} }{G\, \sqrt{N_{pol} \cdot B \cdot T_{\rm obs}}} \cdot \frac{W_{10}}{P - W_{10}}.
\end{equation}
Here, $S_{\rm mean}$ is the mean pulse flux density; $\beta = 1$ is the correction factor for FAST; $S/N$ is the signal-to-noise ratio; $T_{\rm sys}$ is the system temperature in kelvin; $G \approx 16.12$\,K/Jy is the telescope gain \citep{Jiang20}; $N_{pol} = 2$ is the number of polarizations summed; $B = 400$\,MHz is the effective bandwidth; $T_{\rm obs} = 3600$\,s is the integration time; $W_{10}$ is the pulse width at 10\% of the peak; and $P$ is the pulsar period.

The system temperature $T_{\rm sys}$ as a function of zenith angle $\theta_{\rm ZA}$ is calculated using the empirical formula:
\begin{equation}
	T_{\rm sys} = P_{0}\, \arctan\left(\sqrt{1 + \theta_{\rm ZA}^{n}} - P_{1}\right) + P_{2},
\end{equation}
where $\theta_{\rm ZA}$ is the zenith angle in degrees (ranging from $0^\circ$ to $40^\circ$), and $P_0$, $P_1$, $P_2$, and $n$ are the fitting parameters specific to each of the 19 beams of the FAST receiver system across different frequencies. In our observations, we used beam M01 with a central frequency of 1250\,MHz. The corresponding parameter values from Table 4 of \citet{Jiang20} are $P_0 = 4.35$, $P_1 = 6.88$, $P_2 = 25.54$, and $n = 1.21$. Finally, the mean flux density of the pulsar was measured to be 72.7\,$\mu$Jy when we consider a zenith angle of $\theta_{\rm ZA} = 20^{\circ}$.

  \begin{figure}
	\centering
	\includegraphics[width=\columnwidth, angle=0]{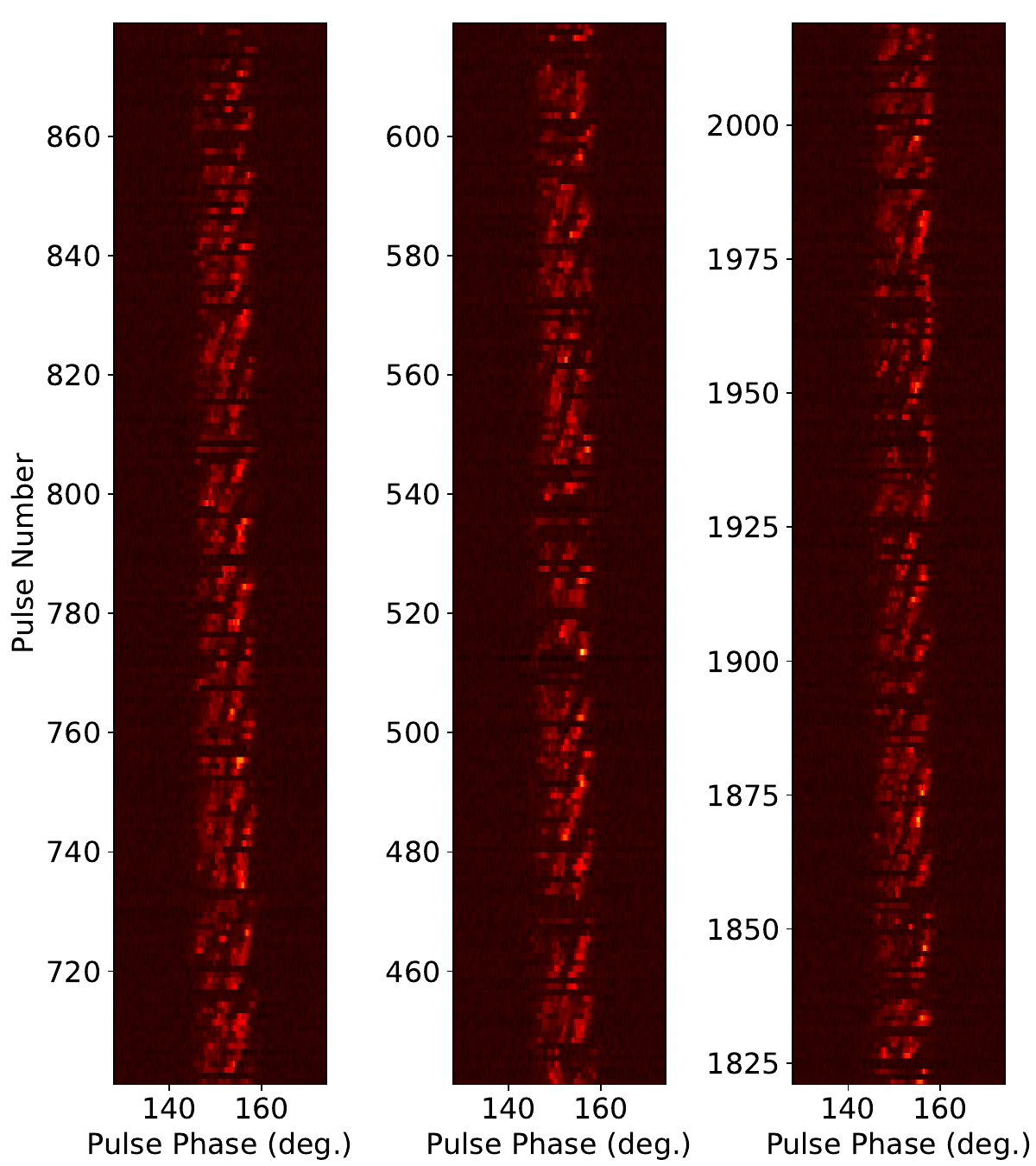}
	\caption{Pulse stacks of PSR J2129+4119 observed on 2024 September 29, showing three distinct behaviors: steady drifting (left), intermittent drifting with breaks (middle), and beat-like modulation (right).}
	\label{stack1}
\end{figure}

  \section{The pulse emission sequences}\label{sec:analysis_result}
  
  Figure~\ref{stack1} shows three consecutive sections of the pulse stack of PSR J2129+4119 observed on 2024 September 29. Drifting subpulses are visible as diagonal bands of emission. In the left panel, the drift bands appear regular and persistent over a long duration. In contrast, the middle panel shows short-lived drift sequences with interruptions, while the right panel exhibits beat-like modulation, possibly due to competing drift modes or amplitude variations. A particularly notable feature is the slight curvature of the drift bands, which becomes steeper toward the latter part of the profile, corresponding to the trailing and strongest component. 
  In addition to subpulse drifting, the single-pulse sequences of PSR J2129+4119 display a variety of emission states, including regular pulses and nulling. We also identify a distinct population of weak pulses, which are too faint to be considered regular emission but do not meet the criteria for nulls. At the other extreme, a small number of bright pulses are observed, with peak intensities exceeding ten times that of the average pulse profile.

\begin{figure}
	\centering
	\includegraphics[width=\columnwidth, angle=0]{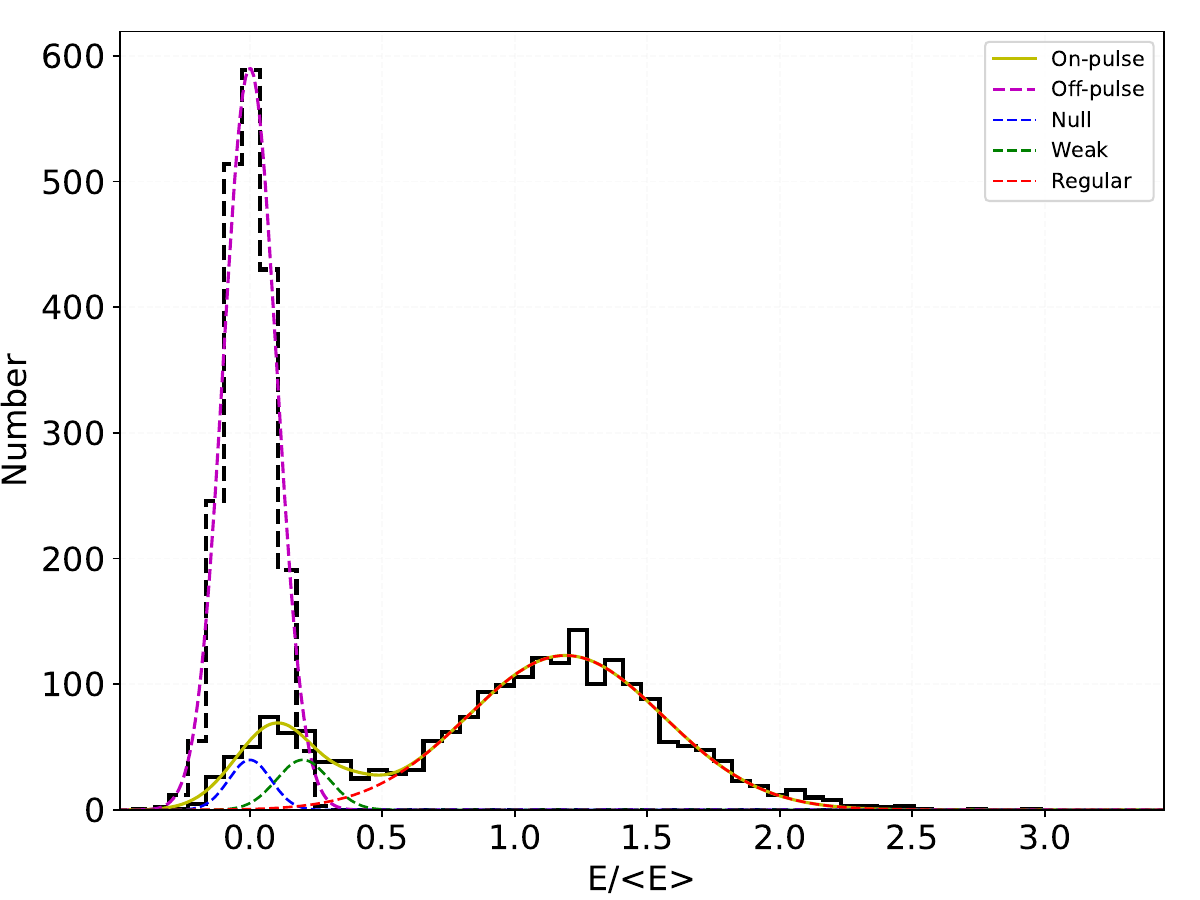}
	\caption{Pulse energy distributions for PSR J2129+4119. The magenta curve represents the Gaussian fit to the off-pulse distribution, while the yellow curve shows the composite fit to the on-pulse distribution using three Gaussian components. The blue, green, and red Gaussian components correspond to the null, weak, and regular emission states, respectively. The $x$-axis represents pulse energy normalized by the mean on-pulse energy.}
	\label{energy}
\end{figure}

 \subsection{Identification of emission states}\label{Identification}

During our observation of PSR J2129+4119, a total of 2090 consecutive pulses were recorded. The pulse sequence shows intermittent nulls and clear variability in emission intensity. To systematically classify the different emission states, we analyzed the pulse energy distributions derived from the on-pulse and off-pulse windows.
The on-pulse window was defined as the phase range where the average intensity exceeds $3\sigma_{\rm off}$, where $\sigma_{\rm off}$ is the standard deviation of the baseline noise measured from the off-pulse region. For each pulse, the on-pulse energy was calculated by summing the intensities within the on-pulse phase bins after subtracting the baseline level. The off-pulse energy was similarly computed using an equal number of bins from the off-pulse region. This method follows the approach of \citet{Ritchings76}.

Figure~\ref{energy} presents the normalized pulse energy distributions. The off-pulse distribution, shown by the dotted black histogram, follows a Gaussian centered at zero and is fitted with a magenta curve, representing pure noise. In contrast, the on-pulse energy distribution (solid black histogram) exhibits a multimodal structure. It is modeled using a combination of three Gaussian components: a blue component centered near zero corresponding to null pulses, a green component at slightly higher energies representing weak pulses, and a red component peaking near $E/\langle E \rangle \approx 1.2$ corresponding to regular emission.
The sum of these three Gaussian components forms the overall fit to the on-pulse distribution, shown in yellow. This multi-component fitting approach enables the statistical separation of emission states and supports a detailed analysis of emission variability in PSR J2129+4119. 
The Gaussian function used to model each emission component is: 

\begin{equation}
	P(E) = \frac{\alpha}{\sqrt{2\pi} \, \sigma} \exp\left[ -\frac{(E - \mu)^2}{2\sigma^2} \right],
\end{equation}\\
where \( E \) denotes the pulse energy, \( \alpha \) is a normalization constant (amplitude), \( \mu \) represents the mean energy level of the component, and \( \sigma \) corresponds to the standard deviation. The fitted parameters for the three Gaussian components associated with the emission states are presented in Table~\ref{table:em1}.

\begin{table}
	\centering
	\caption{Parameters of the Gaussian components used to fit the pulse energy distribution of PSR~J2129+4119, corresponding to the null, weak, and regular emission states shown in Figure~\ref{energy}.}
	\setlength{\tabcolsep}{16pt}
	\begin{tabular}{cccc}
		\hline\hline
		Parameter & Null & Weak & Regular \\
		\hline 
		$\alpha$ & 37.6 & 38.2 & 123 \\
		$\mu$    & 0.00 & 0.20 & 1.19 \\
		$\sigma$ & 0.08 & 0.10 & 0.35 \\
		\hline
	\end{tabular}
	
\vspace{2mm}
	\begin{minipage}{\columnwidth}
		\small \textbf{Note.} $\alpha$ is the amplitude, $\mu$ is the mean, and $\sigma$ is the standard deviation of each Gaussian component.
	\end{minipage}
	\label{table:em1}
\end{table}

\begin{table}
	\setlength{\tabcolsep}{12 pt}  
	\setlength\extrarowheight{3pt}
	\centering
	\caption{Summary of detected null (N), weak (W), and regular (R) pulses in PSR~J2129+4119 from our observation.}
	\begin{tabular}{cc c c}
		\hline\hline
		Emission & Pulses & Abundance & Duration\\
		&  & (\%) & (s) \\
		\hline
		N & 170 & $8.13$ & 292.82\\
		W & 56 & $2.68$ & 96.46 \\
		R & 1864 & $89.19$ & 3210.72 \\
		\hline
	\end{tabular}
	
	\vspace{2mm}
	\begin{minipage}{\columnwidth}
		\small \textbf{Note.} The total time spent in each emission state is shown in seconds.
	\end{minipage}
	\label{table:em2}
\end{table}

Building on this classification, we quantified the nulling behavior of PSR J2129+4119 by estimating the null fraction (NF), defined as the ratio of null pulses to the total number of observed pulses. To identify null pulses from background noise and low-level emission, we used a histogram subtraction technique described by \citet{Ritchings76}. The off-pulse energy distribution was scaled and subtracted from the on-pulse distribution to remove baseline noise contributions. The scaling factor was chosen such that the net count in bins with energy $E < 0$ bins was zero \citep{Wang07}, effectively preventing false null identifications due to noise fluctuations. Pulses falling within the energy range associated with the null component of the fitted model were classified as nulls. The uncertainty in NF was estimated using the expression $\sqrt{n_p} / N$, where $n_p$ is the number of identified null pulses and $N$ is the total number of pulses analyzed \citep{Wang07}. Our analysis yields a null fraction of $8.13\% \pm 0.51\%$ for PSR~J2129+4119.

\begin{figure}
	\centering
	\includegraphics[width=\columnwidth, angle=0]{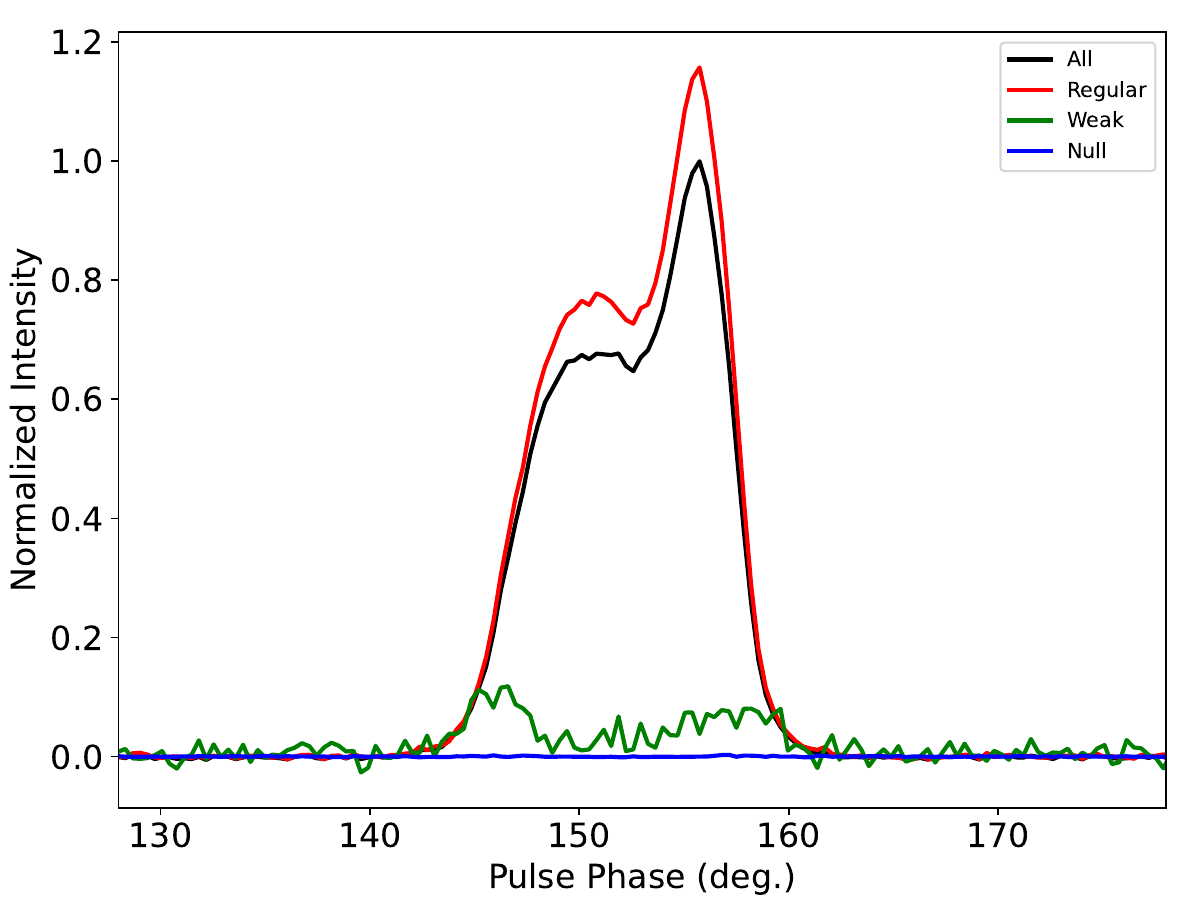}
	\caption{Integrated pulse profiles of PSR J2129+4119 for regular (red), weak (green), and null (blue) pulses. The regular and weak profiles are normalized to the integrated profile of the entire observation (black).}
	\label{Subsequences}
\end{figure}

To further distinguish between null, weak, and regular pulses, we adopted a statistical thresholding method proposed by \citet{Bhattacharyya10} and later refined by \citet{HMT2025}. In this approach, an energy threshold for each single pulse is determined using the relation $\xi_{\rm on} = \sqrt{N_{\rm on}}\,\sigma_{\rm off}$, where $N_{\rm on}$ is the number of bins in the on-pulse window and $\sigma\textit{}_{\rm off}$ is the root mean square (RMS) of the off-pulse region calculated over the same number of bins. Pulses with energy less than or equal to $3\times\xi_{\rm on}$ are classified as nulls; those with energy between $3\times\xi_{\rm on}$ and $5\times\xi_{\rm on}$ are categorized as weak pulses; and pulses with energy greater than or equal to $5\times\xi_{\rm on}$ are identified as regular emission.
To quantitatively assess whether the weak-pulse component represents a distinct emission state rather than a low-energy tail of the regular emission, we performed model selection using the Akaike Information Criterion (AIC; \citealt{Akaike1974}) and the Bayesian Information Criterion (BIC; \citealt{Schwarz1978}). We compared a two-component (null + burst) and a three-component (null + weak + regular) Gaussian mixture model (GMM; \citealt{Kaplan2018}), with the null component tied to the independently fitted OFF-pulse distribution. The results yield $\Delta$AIC~$=-145.3$ and $\Delta$BIC~$=-128.4$ (negative values favor the three-component model), indicating strong statistical support for the inclusion of the weak-pulse component. This confirms that the weak pulses form a distinct population between the null and regular states.

\begin{figure}
	\centering
	\includegraphics[width=\columnwidth, angle=0]{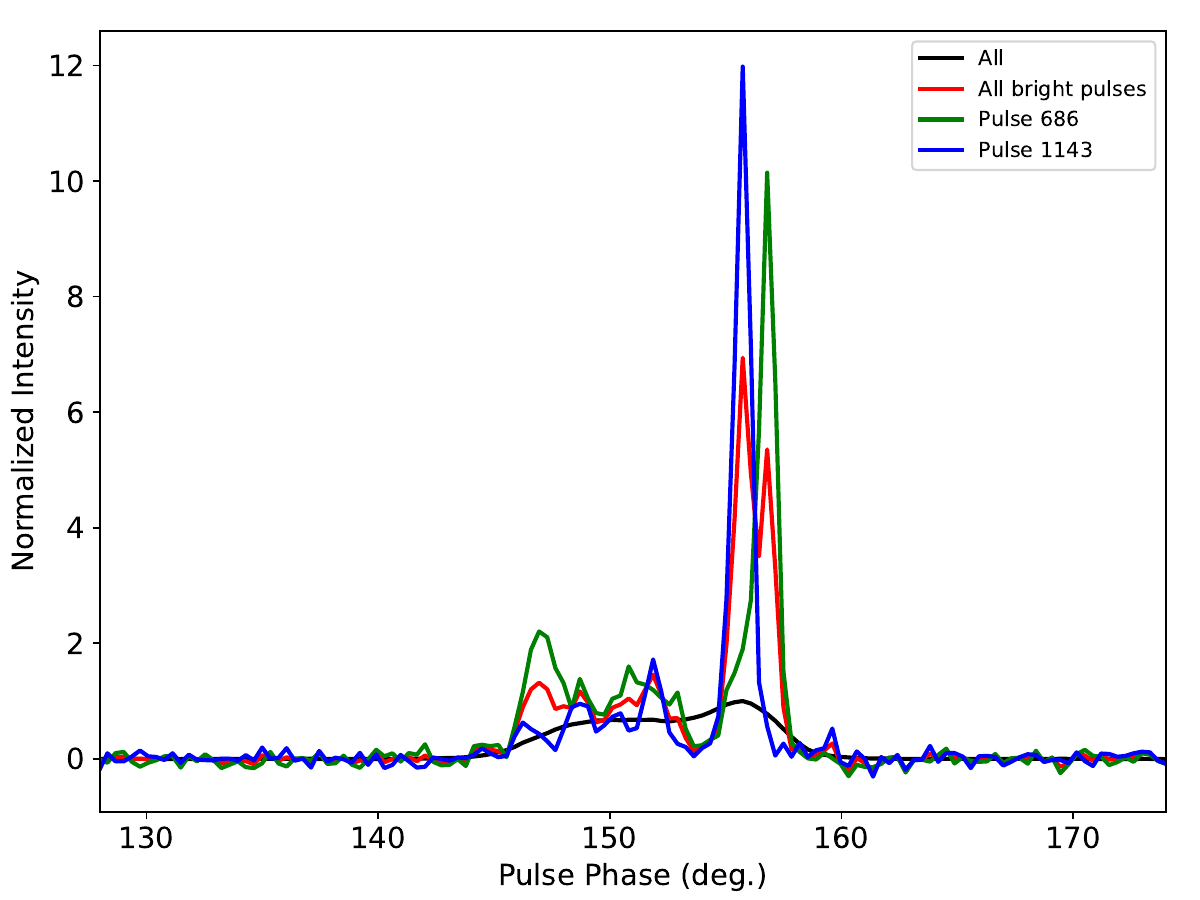}
\caption{Bright pulse profile at pulse numbers 686 (green) and 1143 (blue) compared with the integrated profile from the entire observation (black). The red line represents the integrated profile of all bright pulses detected.}
	\label{brightpulse}
\end{figure}

\subsection{Pulse profiles in different emission states}\label{pulse-profiles-across}

Figure~\ref{Subsequences} shows the integrated pulse profiles for the null, weak, and regular emission states, plotted in blue, green, and red, respectively. The profile from the entire observation is shown in black for comparison. Both the regular and weak profiles are normalized to the entire observation profile.
The regular pulse profile closely resembles the overall profile, which is expected given that regular pulses dominate the dataset, comprising approximately 89.19\% of all detected pulses. Two bright pulses are identified within the regular emission category, as illustrated in Figure~\ref{brightpulse}. These are occasional, individual pulses whose peak intensities significantly exceed those of the average regular pulses. Specifically, we define bright pulses as those with peak intensities exceeding ten times the mean intensity of the integrated profile. The pulse exhibits pronounced asymmetry in shape between the leading and trailing components, with the trailing side dominating in intensity. Additionally, the profile reveals fine-scale structure within the main peak, suggesting the presence of microstructure emission. These characteristics indicate multi-component, possibly short-timescale emission processes that are not evident in the average pulse profile. Additional observational evidence is presented in Section~\ref{sec:microstructure}. This behavior is consistent with our earlier findings \citep{HMT2025}, where we reported similar single-pulse events in the long-period pulsar PSR~J1900--0134, observed using the FAST telescope.

In contrast, the weak pulse profile displays a distinct morphology characterized by a nearly symmetric structure, though the leading component is slightly more intense than the trailing one. A visible bridge connects the two components, indicating a smoother temporal transition and enhanced pulse-to-pulse variability. These characteristics are subdued or absent in the regular emission profile, which is more asymmetric and dominated by the trailing component.
Additionally, null pulses appear slightly more frequently than weak pulses, based on their respective proportions in the total pulse sequence.

A comparison of pulse widths reveals that, at the 10\% ($W_{10}$) and 50\% ($W_{50}$) intensity levels, the regular pulse profiles are approximately 2.63\% and 3.70\% wider than the full observation profile, respectively. Furthermore, the peak intensity of the regular profile is about 13.62\% higher than that of the entire observation profile, and approximately 89.82\% higher than the peak of the weak pulse profile. These distinctions highlight the necessity of analyzing each emission state independently better to understand the underlying emission mechanisms and their temporal variability. The occurrence statistics for the null, weak, and regular pulses are summarized in Table~\ref{table:em2}.

\subsection{The average polarimetric profiles}\label{polarization}

\begin{figure}
	\centering
	\includegraphics[width=\columnwidth, angle=0]{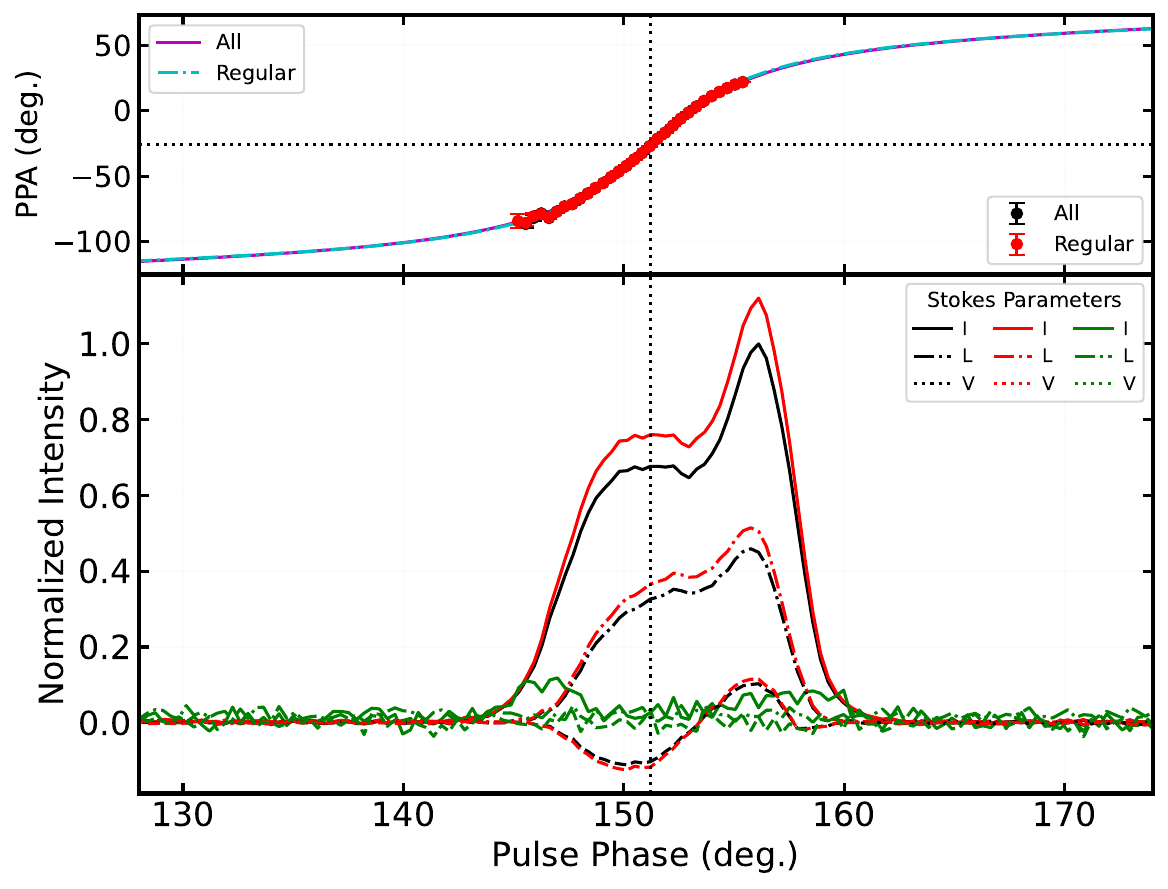}
	\caption{
		Polarization properties of PSR~J2129+4119 for three cases: regular (red), weak (green), and the entire observation (black).
		Top panel: Position angle (PPA) as a function of pulse phase, with best-fit RVM curves for the full observation (magenta) and regular mode (cyan). The intersection of the vertical and horizontal dashed lines marks the point of steepest PPA gradient. 
		Bottom panel: Normalized Stokes parameters of the total intensity (\(I\)), linear polarization (\(L \)), and circular polarization (\(V\)) are plotted across pulse phase for all three datasets. All values are normalized to the peak intensity of the entire observation profile.
	}
	\label{RVM_fit}
\end{figure}

Figure~\ref{RVM_fit} presents the polarization position angle (PPA) swing and normalized Stokes parameter profiles of PSR~J2129+4119, shown for the entire observation (black), the regular mode (red), and the weak mode (green). In the top panel, PPA is plotted as a function of pulse phase, overlaid with the best-fit Rotating Vector Model \citep[RVM;][]{RadhakrishnanCooke1969} curves for the entire observation (magenta) and the regular mode (cyan). The PPA distributions from the entire and regular mode datasets are nearly identical and well-described by the RVM fits. In the leading component (Pulse phase $\lesssim 148^\circ$), the PPA points are sparsely detected and exhibit larger uncertainties, likely due to reduced linear polarization or lower signal-to-noise in that phase range. No significant PPA points were detected in the weak mode. Consequently, only the full and regular mode PPA data were used in the RVM fitting. 

The bottom panel displays the normalized Stokes parameters: total intensity (\(I\)), linear polarization (\(L = \sqrt{Q^2 + U^2}\)), and circular polarization (\(V\)), all scaled by the peak intensity of the entire observation. The regular mode exhibits slightly higher total intensity and linear polarization compared to the entire observation profile, while its circular polarization curves are nearly identical. The PPA is calculated as \(\psi = \tan^{-1} \left( \frac{U}{Q} \right)\), where \(U\) and \(Q\) are the Stokes parameters representing the orthogonal components of linear polarization.

\begin{table*}[t]
    \centering
    \caption{Best-fit RVM parameters for PSR~J2129+4119 from the full data set (`All') and the regular emission mode (`Regular').}
    \setlength{\tabcolsep}{25 pt} 
    \renewcommand{\arraystretch}{1.05}
    \begin{tabular}{ccccc}
        \hline\hline
        Emission & $\alpha$ (°) & $\beta$ & $\psi_0$ & $\phi_0$ \\
        \hline
        All     & $44.08 \pm 17.9$ & $-2.69 \pm 1.98$ & $-26.07 \pm 2.13$ & $151.19 \pm 0.44$ \\
        Regular & $50.83 \pm 13.15$ & $-2.84 \pm 1.55$  & $-26.04 \pm 1.72$ & $151.21 \pm 0.36$ \\
        \hline
    \end{tabular}
    \label{tab:rvm_fit}
\end{table*}

The emission geometry of PSR~J2129+4119 is modeled using the RVM \citep{RadhakrishnanCooke1969}, where the position angle, \(\psi\), of the linearly polarized emission is expressed as a function of pulse phase, \(\phi\), using the formulation of \citet{Everett01}:
\begin{equation}
	\tan(\psi - \psi_0) = \frac{\sin\alpha \sin(\phi - \phi_0)}{\sin\zeta \cos\alpha - \cos\zeta \sin\alpha \cos(\phi - \phi_0)},
\end{equation}
where \(\alpha\) is the magnetic inclination angle, \(\zeta = \alpha + \beta\) is the observer’s viewing angle, \(\phi_0\) is the pulse phase corresponding to the closest approach of the line of sight to the magnetic axis, and \(\psi_0\) is the PPA at that fiducial phase. Following \citet{Everett01}, we modified the original RVM equation to match the astronomical convention \citep[IAU/IEEE;][]{vanStraten10} in which the position angle increases counterclockwise on the sky. 
To constrain the geometric parameters \(\alpha\), \(\beta\), \(\psi_0\), and \(\phi_0\), we employed a Markov Chain Monte Carlo (MCMC) method\footnote{https://github.com/dfm/corner.py} \citep{emcee, corner}, allowing us to obtain posterior distributions and credible intervals for each parameter associated with the pulsar's emission geometry.

\begin{table}[t]
    \centering
    \caption{Mean polarization ratios and their uncertainties for the entire observation, regular, and weak emission profiles of PSR~J2129+4119.}
    \setlength{\tabcolsep}{9 pt} 
    \renewcommand{\arraystretch}{1.1}
    \begin{tabular}{lccc}
        \hline\hline
        Emission & $\langle L \rangle / \langle I \rangle$ & $\langle V \rangle / \langle L \rangle$ & $\langle |V| \rangle / \langle L \rangle$ \\
        \hline
        All (Black)   & 51.6(11) & $-5.8(3)$  & 31.7(5) \\
        Regular (Red) & 51.5(11) & $-4.6(3)$  & 31.6(5) \\
        Weak (Green)  & 25.4(7) & $-28.3(5)$ & 86.2(31) \\
        \hline
    \end{tabular}

    \vspace{2mm}
    \begin{minipage}{\columnwidth}
        \small \textbf{Note.} All values are expressed in percent. Numbers in parentheses indicate $1\sigma$ uncertainties on the last digits.
    \end{minipage}
    \label{tab:pol_ratios}
\end{table}

\begin{figure}
	\centering
	\includegraphics[width=\columnwidth]{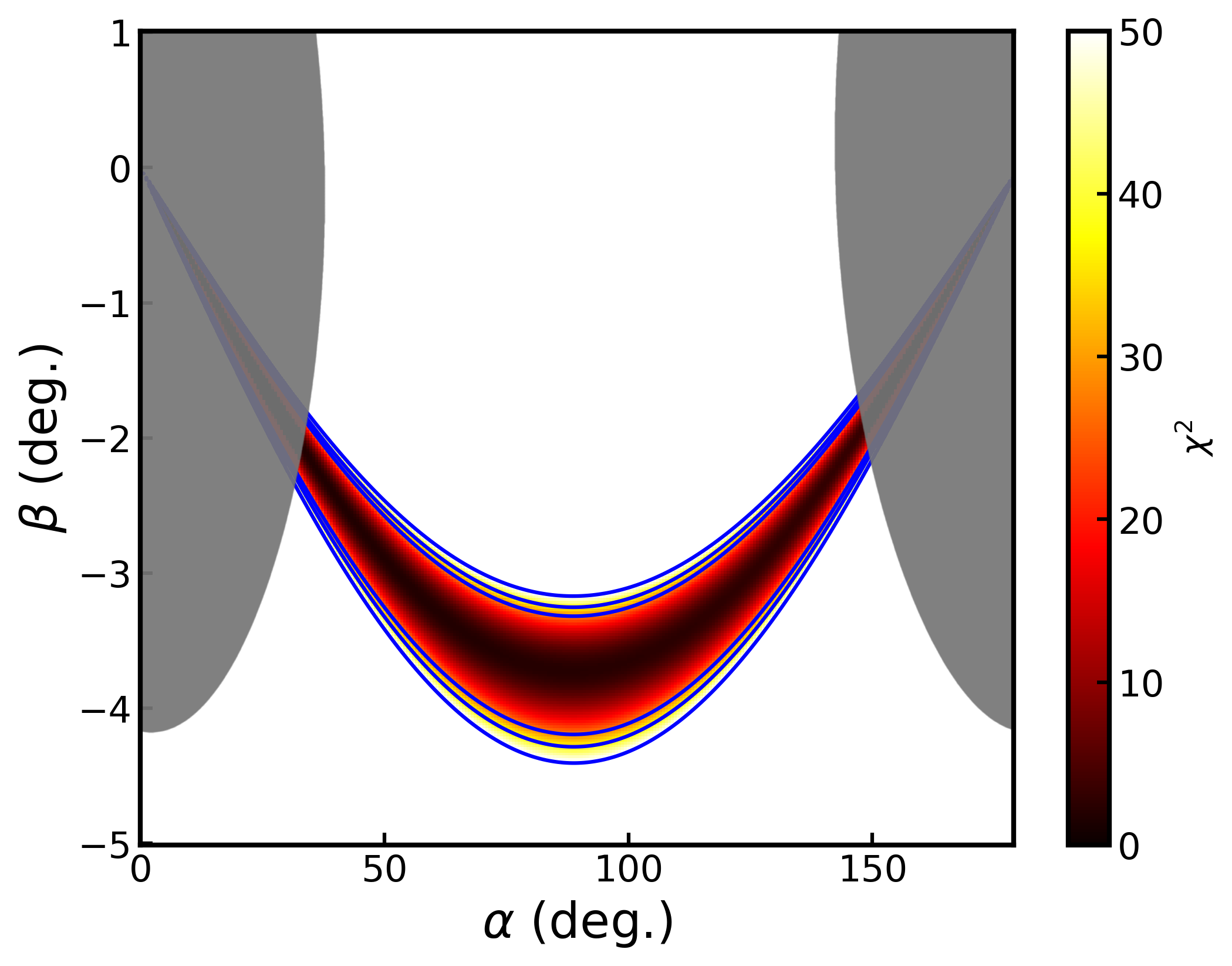}
	\caption{Reduced $\chi^2$ contour map in the $(\alpha, \beta)$ combinations from RVM fitting of the entire observation. The hot scale indicates the fit quality, with darker regions corresponding to lower $\chi^2$ values (better fits). The blue contours represent the 1$\sigma$, 2$\sigma$, and 3$\sigma$ confidence intervals. The banana-shaped region reflects the strong correlation between $\alpha$ and $\beta$, resulting in a poorly constrained $\alpha$. The shaded gray region denotes viewing geometries excluded by the observed pulse width constraint.}
	\label{alpha_beta_contour}
\end{figure}

We performed RVM fitting separately on the PPA data from the full dataset (magenta) and the regular emission mode (cyan). The best-fit parameters for both cases are summarized in Table~\ref{tab:rvm_fit}, with reduced chi-squared values of 9.77 for the full dataset and 12.32 for the regular mode. 
The fitted values of $\phi_0$ and $\psi_0$ are nearly identical for both the full dataset and the regular emission mode, and the PPA swings visually overlap, indicating consistent polarization behavior. However, $\alpha$ is slightly larger in the regular mode ($50^\circ.83$) compared to the full dataset ($44^\circ.08$), and the corresponding $\beta$ values also differ modestly. In both cases, the small $\beta$ implies that the line of sight passes close to the magnetic axis, resulting in a near-tangential traverse across the emission beam, though not exactly tangential, as $\alpha \ne \zeta$.
Despite their similarity, the PPA trajectory in the regular mode is smoother and less scattered, suggesting reduced depolarization and minimal orthogonal mode mixing. This leads to a more reliable RVM fit, supporting the interpretation that the regular mode reflects a more stable emission state. The close alignment of $\phi_0$ with the phase of peak emission in both fits further reinforces the geometrical consistency.

Figure~\ref{alpha_beta_contour} presents the reduced \(\chi^2\) contour map in the \((\alpha, \beta)\) combinations derived from the RVM fitting, revealing a banana-shaped region of high likelihood. While \(\beta\) is relatively well constrained (\(|\beta| \lesssim 4^\circ\)), \(\alpha\) remains poorly determined over a broad range, approximately from \(30^\circ\) to \(150^\circ\). This degeneracy reflects the intrinsic limitations of RVM fitting when the PPA swing spans a limited longitude range or when \(|\beta|\) is small. The darkest region in the plot indicates the best-fit parameter combination. The shaded gray area denotes viewing geometries excluded by the observed pulse width, based on the relationship between the emission beam opening angle and the geometry \citep[e.g.][]{Gil1984, Mitra2011}. A more detailed analysis of these geometric constraints is provided in Section~\ref{sec:discus}.

The inflection point of the RVM curve, corresponding to the steepest gradient (SG) in the PPA swing, occurs at $\phi_0 = 151^{\circ}.19 \pm 0^{\circ}.44$ and $\psi_0 = -26^{\circ}.07 \pm 2^{\circ}.13$, as indicated by the vertical and horizontal gray dashed lines in the upper panel of Figure~\ref{RVM_fit}. The steepest gradient is given by
\begin{equation}
	\left( \frac{d\psi}{d\phi} \right)_{\rm max} = \frac{\sin\alpha}{\sin\beta}.
\end{equation}
Using the best-fit parameters from the full dataset ($\alpha = 44^{\circ}.08 \pm 17^{\circ}.9$, $\beta = -2^{\circ}.69 \pm 1^{\circ}.98$), we derive an SG value of $-14.82 \pm 11.91$. The negative value of $\beta = \zeta - \alpha$ implies the observer's line of sight passes below the magnetic axis, resulting in a positive slope in the PPA curve. This SG point is located close to the center of the pulse profile \citep{Lyne1988} but shows a slight shift toward the leading component. Such displacement has been observed in slow-rotating pulsars and is often attributed to aberration and retardation effects that alter the apparent emission geometry \citep{Blaskiewicz1991, Mitra2011}. 

Previous studies by \citet{Cruces2021} reported that PSR J2129+4119 exhibits a high degree of linear polarization (exceeding 70\%). In our analysis, we also detect a significant level of linear polarization, with a moderate average value of approximately 52\%. The regular emission mode exhibits strong linear polarization and noticeable circular polarization, with the circular component changing sign between the leading (positive) and trailing (negative) components. In contrast, the weak mode displays substantially lower total intensity but maintains comparable levels of linear and circular polarization across a similar pulse phase range. The polarization ratios, summarized in Table~\ref{tab:pol_ratios}, indicate a moderate linear polarization fraction in the regular mode (⟨$L$⟩/⟨$I$⟩). 

The on-pulse region was initially defined in Section~\ref{Identification} from the overall average profile using $I > 3\sigma_I$, where $\sigma_I$ is the rms noise level of the Stokes~$I$ parameter estimated from the off-pulse region. This approach ensures a consistent pulse-phase window for all modes. For the polarization-ratio estimates, the same criterion was applied separately to each mode to exclude bins where the signal was indistinguishable from noise. 

\begin{figure}
	\centering
	\includegraphics[width=\columnwidth]{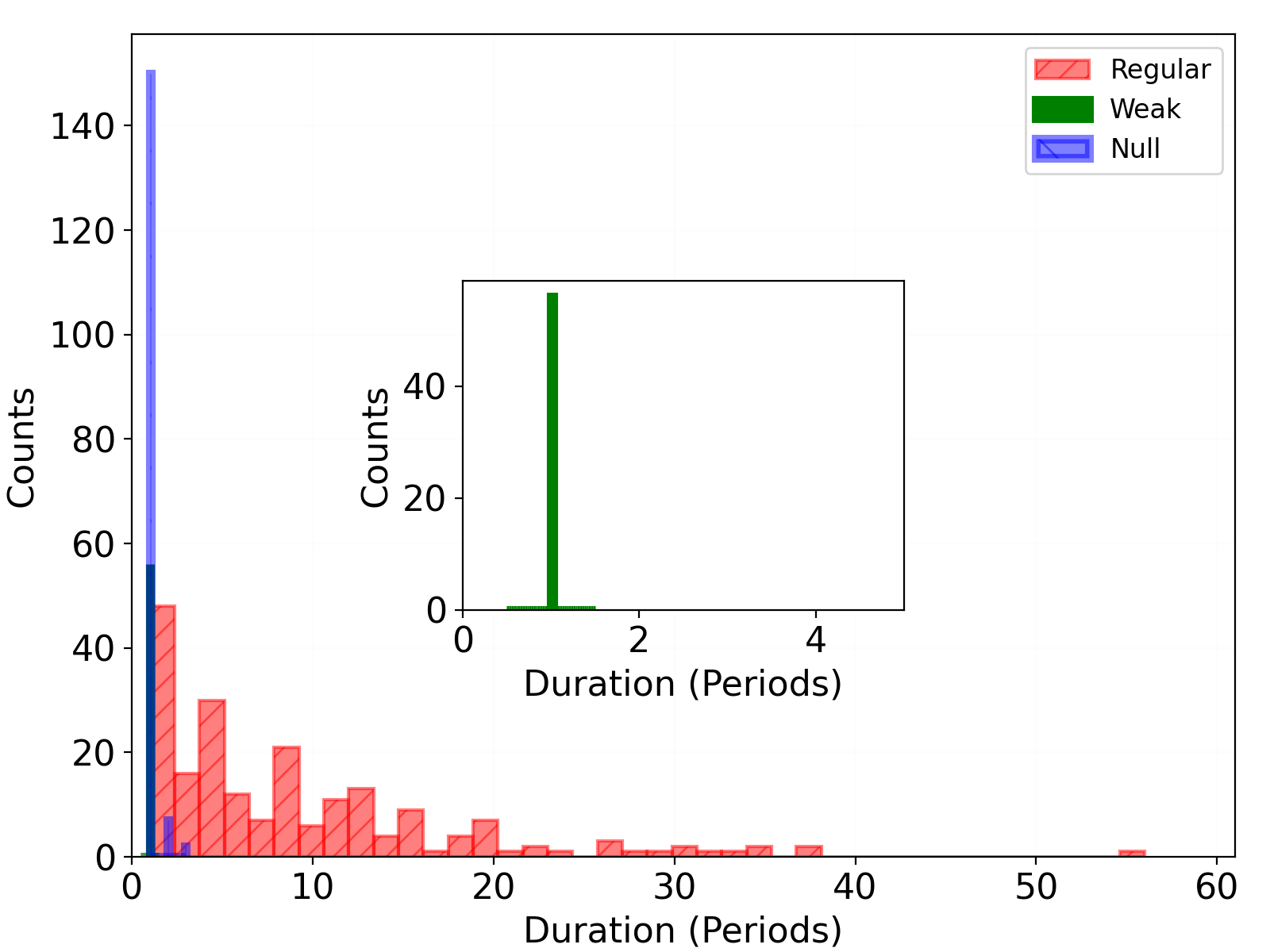}
	\caption{Histograms of the durations of null (blue), weak (green), and regular (red) emission states in PSR~J2129+4119. The inset shows the duration distribution for weak pulses.} 
\label{Pulsedurations}
\end{figure}

The effective noise in the linear polarization was taken as 
$\sigma_L = 0.5(\sigma_Q + \sigma_U)$. 
The linear polarization was corrected for positive noise bias using the standard Ricean debiasing relation 
\begin{equation}
L' = \sqrt{Q^2 + U^2 - \sigma_L^2}
\end{equation}
(\citealt{WardleKronberg1974}), and bins with $|V| < 3\sigma_V$ were excluded to suppress spurious noise contributions. The resulting polarization ratios are listed in Table~\ref{tab:pol_ratios}. 
The strong reduction in linear polarization for the weak mode is consistent with its low signal-to-noise nature. The apparently high $\langle |V| \rangle / \langle L \rangle$ value arises because the debiased linear polarization becomes very small while the Stokes~$V$ fluctuations remain comparable to the off-pulse noise; this ratio should therefore be regarded as an upper limit rather than evidence for circular-polarization dominance.

\begin{figure}
	\centering
	\includegraphics[width=\columnwidth, angle=0]{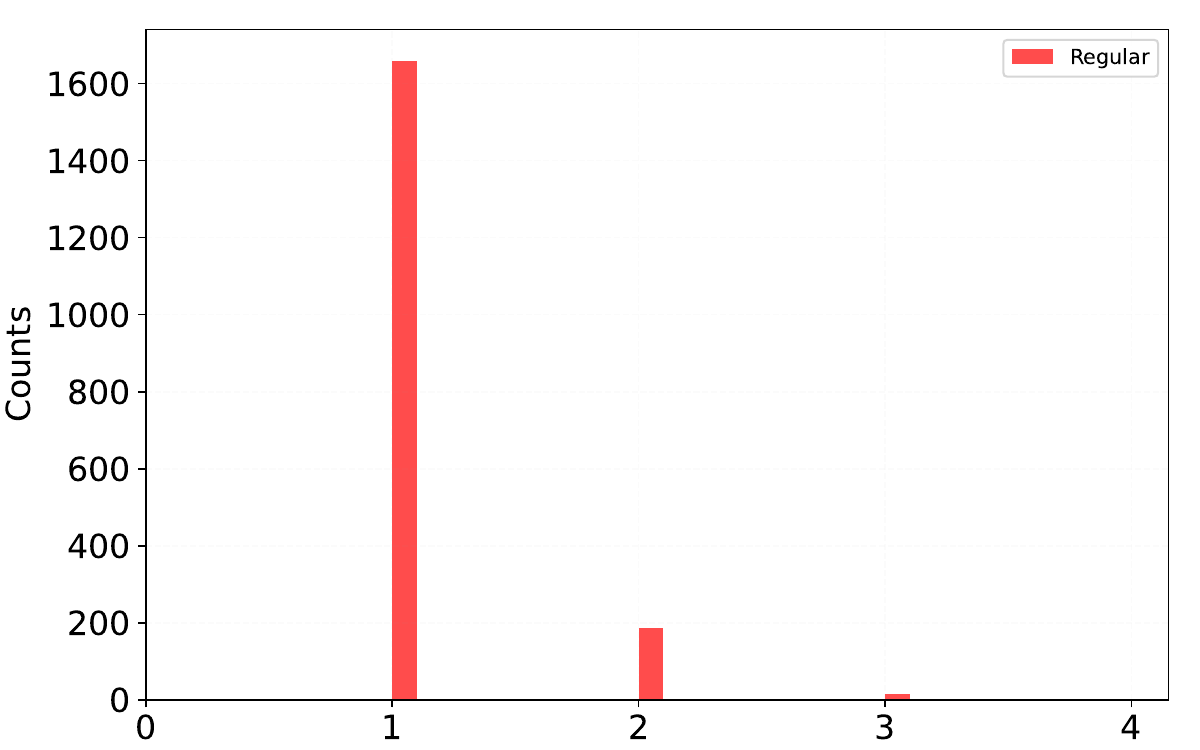}
	\includegraphics[width=\columnwidth, angle=0]{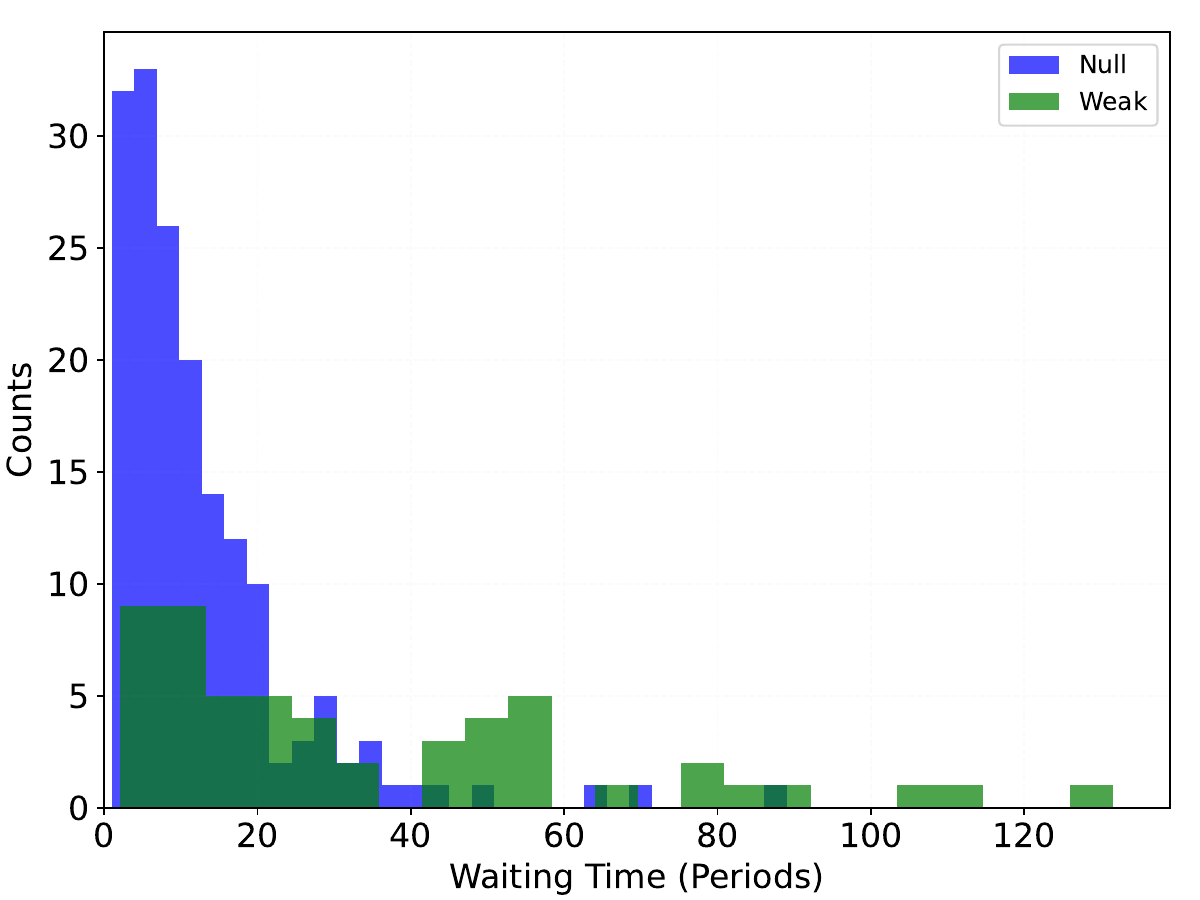}
	\caption{Waiting-time distributions for the null (blue) and weak (green) emission states (bottom panel) and for the regular emission state (red, top panel) of PSR~J2129+4119.}
	\label{WaitingTimes}
\end{figure}

\subsection{Pulse durations and waiting times}\label{waiting-times}

Figure~\ref{Pulsedurations} shows the duration histograms for the three emission states: regular (red), null (blue), and weak (green, shown in the inset for clarity). The distribution of regular durations spans a broad range, extending up to 56 pulse periods, with a gradual decline in frequency and a long tail. The majority of regular sequences last between 1 and 20 pulse periods, with a mean duration of approximately 28.5 periods.

Null durations, shown in blue, are more concentrated and sharply peaked at shorter timescales, with most nulls lasting for only one or two periods, and none exceeding three pulse periods. Weak pulses, represented in green, occur exclusively as isolated single-pulse events. This indicates that the weak emission state does not persist over multiple consecutive periods, in contrast to both regular and null states.

Figure~\ref{WaitingTimes} presents the waiting-time distributions for the null, weak, and regular emission states, illustrating how long the pulsar remains in a given state before transitioning to another. The null and weak state distributions are shown in the bottom panel in blue and green, respectively, while the regular state distribution is displayed in the top panel.
For the null states, the waiting times extend up to approximately 88 pulse periods, with a steadily decreasing trend in frequency indicating that shorter null sequences are more common. However, occasional longer null intervals do occur. 

A similar distribution is observed for weak pulses, with waiting times also reaching up to 130 periods; however, the overall counts are noticeably lower.
In contrast, the waiting-time distribution for regular emission is sharply peaked at the shortest durations, indicating that transitions from other states into the regular state typically occur after very short intervals. Most regular sequences begin within 1–2 pulse periods, and the frequency drops off rapidly for longer waiting times. This behavior illustrates the dominant and persistent nature of the regular state in PSR~J2129+4119.

\subsection{Emission variability near nulls}

\begin{figure}
	\centering
	\includegraphics[width=\columnwidth, angle=0]{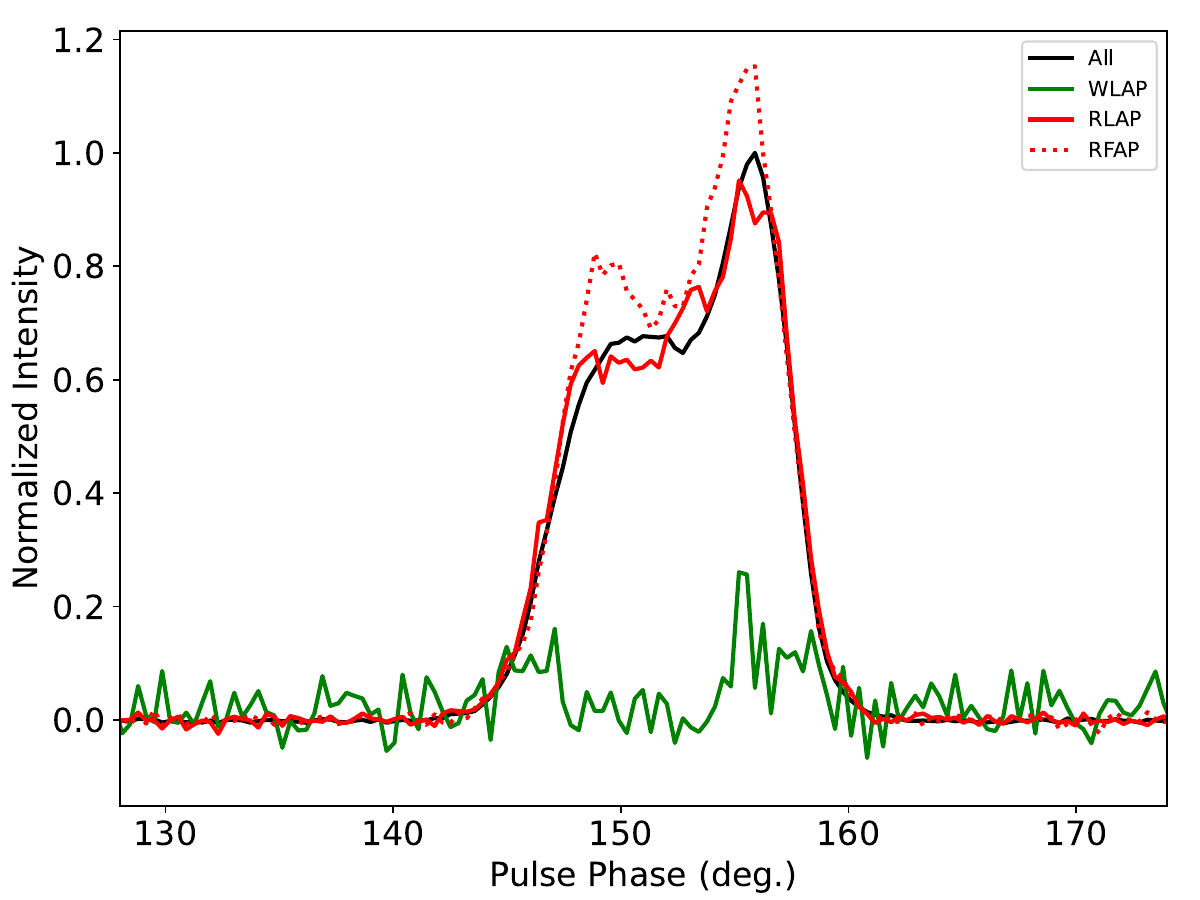}
	\caption{Integrated pulse profiles of weak and regular pulses occurring immediately before and after nulls in PSR J2129+4119 (see main text for definitions). All profiles are normalized to the integrated profile of the entire observation (black).}
	\label{Duration_both}
\end{figure}

To investigate how emission behavior changes near nulls, we categorized single pulses occurring immediately before and after nulls into four types based on their intensity and temporal position relative to the null sequence. Specifically, we define regular pulses before and after nulls as RLAP and RFAP, respectively, and weak pulses in the same positions as WLAP and WFAP. Table~\ref{table:em3} summarizes the number of pulses in each category, while the corresponding integrated profiles are shown in Figure~\ref{Duration_both}.
This classification enables a detailed evaluation of emission transitions and symmetry around nulls. 

	\begin{table}
		\setlength{\tabcolsep}{23pt}
		\centering
		\caption{Statistics for single-pulse emission before and after nulls.}
		\begin{tabular}{ccc}
			\hline
			\hline
			Emission & Before & After \\
			\hline
			Regular       & 151 & 158 \\
			Weak       & 8   & --   \\
			Bright   & --  & --  \\
			\hline
		\end{tabular}
		\label{table:em3}
	\end{table}
    
\begin{figure*}
	\centering
	\includegraphics[width=2.1\columnwidth, angle=0]{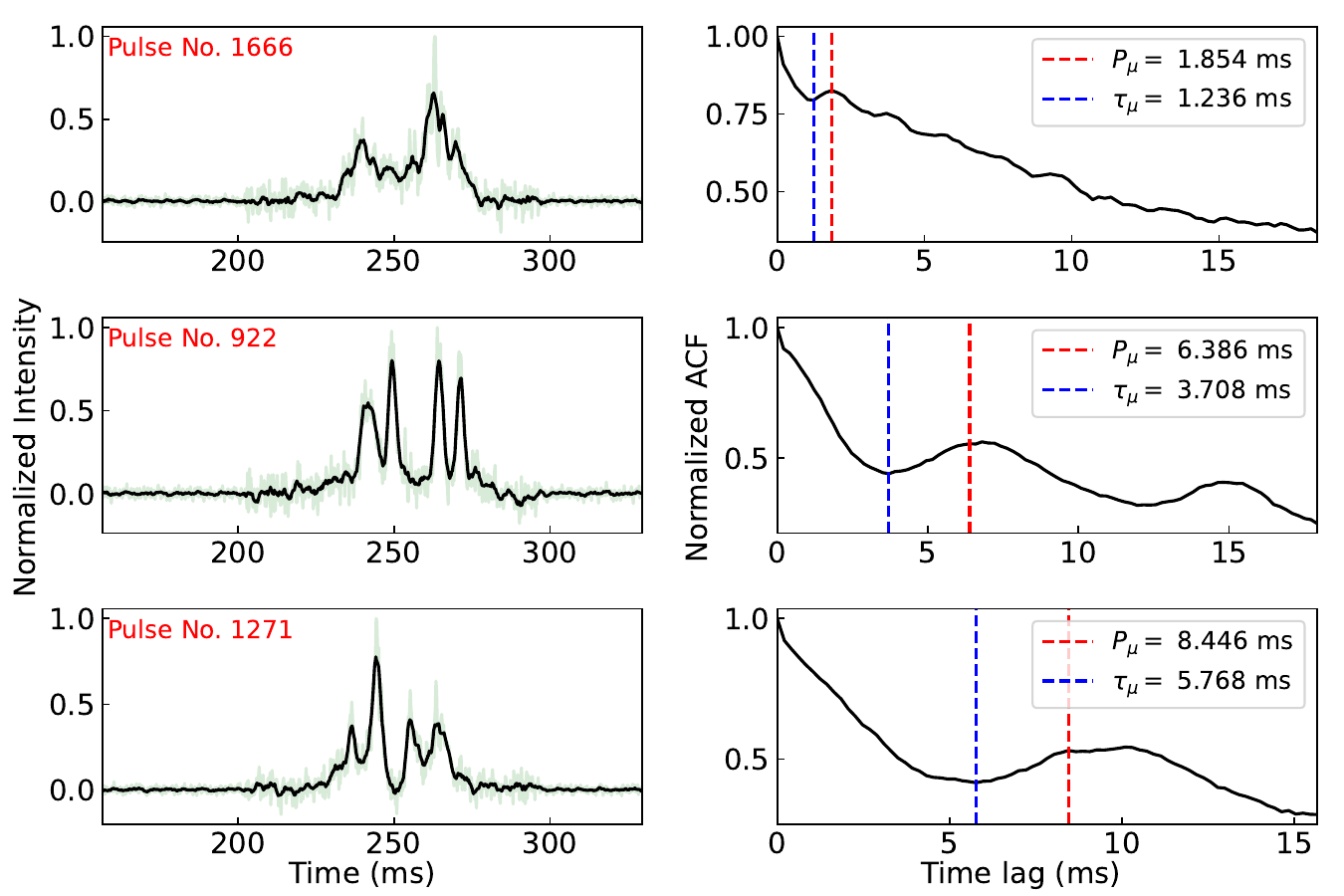}
	\caption{Example of quasiperiodic microstructure pulses detected from PSR~J2129+4119. Left: pulse profiles. Right: corresponding ACFs at 49.25~$\mu$s resolution for pulse numbers 1666 (top), 922 (middle),  and 1271 (bottom). Blue and red dotted lines mark the microstructure width ($t_{\mu}$) and quasiperiodicity ($P_{\mu}$), respectively.}
	\label{Quasi-periodic}
\end{figure*}

The regular pulses (RLAP and RFAP) occur in nearly equal numbers, with 151 before and 158 after nulls. Their integrated profiles, however, reveal subtle but clear asymmetries. RFAP shows enhanced intensity in the trailing component compared to RLAP, and its overall peak intensity is 21.15\% higher. In contrast, RLAP displays a lower peak, particularly in the leading component. This asymmetry suggests that regular emission tends to strengthen immediately after nulls, possibly due to gradual magnetospheric reactivation. Notably, no bright pulses occur immediately before or after nulls; they are observed only between regular pulses.

Weak pulses exhibit a different behavior. Only eight WLAP events are detected before nulls, and no WFAP pulses occur after nulls. The WLAP profile is distinctly asymmetric: the trailing component maintains comparable strength, whereas the leading component is markedly weaker. These patterns suggest that weak emission around nulls is less stable and can vary substantially in both shape and strength across the null boundary.
Overall, these observations suggest that PSR~J2129+4119 exhibits asymmetric emission behavior around nulls. Regular emission recovers more strongly following a null, while weak emission shows fewer events and a more irregular profile structure-potentially reflecting differences in post-null magnetospheric conditions.

\subsection{Quasi-periodic microstructure}\label{sec:microstructure}

Pulsar radio emission often exhibits rapid intensity fluctuations on timescales ranging from several microseconds to a few hundred microseconds, a phenomenon known as microstructure \citep{Drake68}. These microstructures, typically superposed on individual subpulses, are believed to be closely linked to the emission mechanism and magnetospheric processes \citep{Lange1998, Popov2002}. High-time-resolution observations have enabled the detection of such fine temporal features in several pulsars.

In our analysis of PSR~J2129+4119, the integrated total intensity profile in Figure~\ref{Subsequences} and the average polarization profiles shown in Figure~\ref{RVM_fit} exhibit no clear evidence of microstructure. This is consistent with previous findings that the averaging process tends to smear out short-timescale structures present in individual pulses.
To investigate the presence of microstructure in PSR~J2129+4119, we utilized single-pulse data obtained from FAST, which offers both high sensitivity and high-time resolution. This pulsar had not been previously studied for microstructure emission. To ensure accurate detection and to maintain a high signal-to-noise ratio (S/N), we restricted our analysis to the subset of regular pulses, which comprise approximately 89\% of the observed data.

\begin{figure*}
	\centering
	\includegraphics[width=\columnwidth, angle=0]{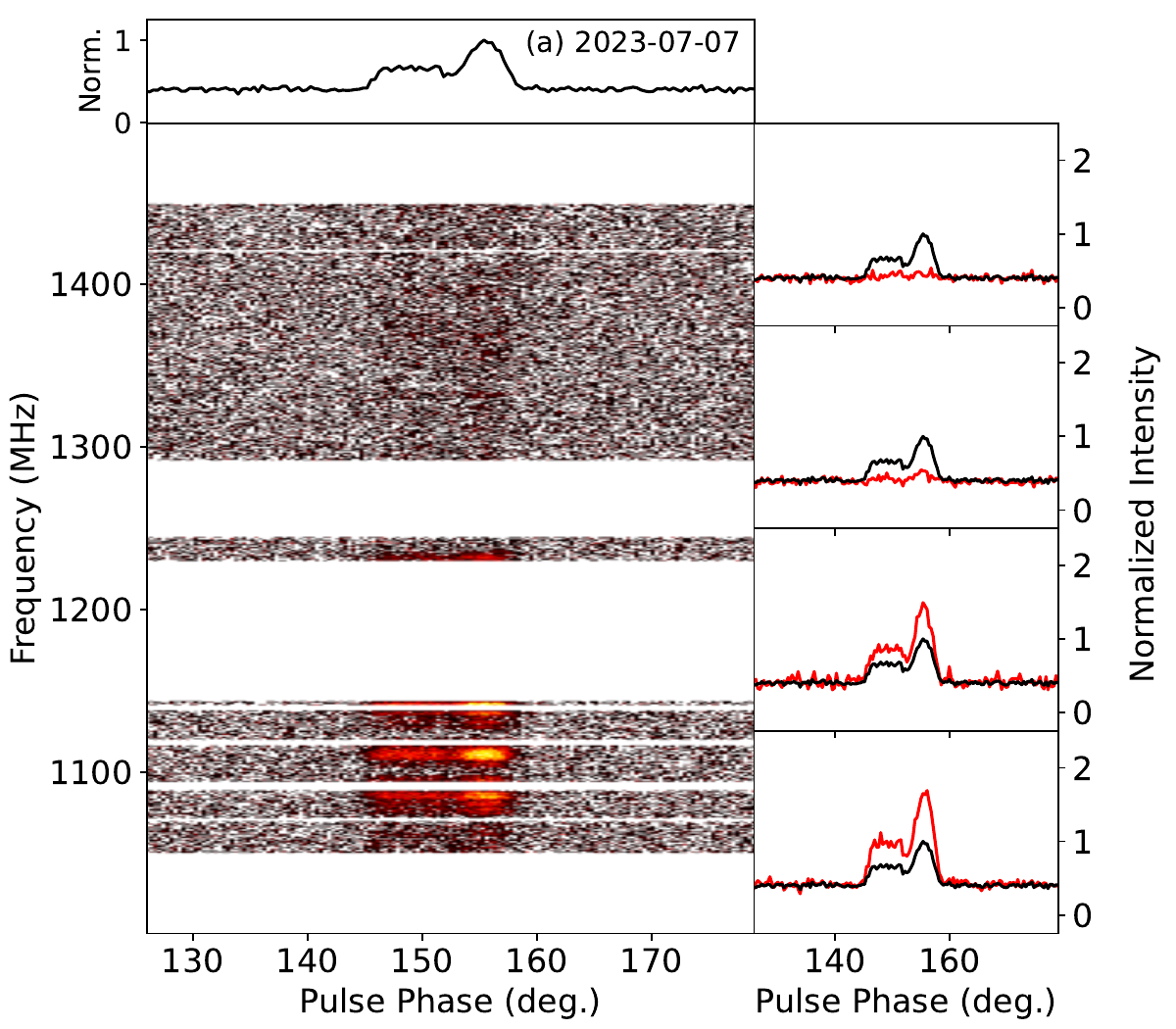}
	\includegraphics[width=\columnwidth, angle=0]{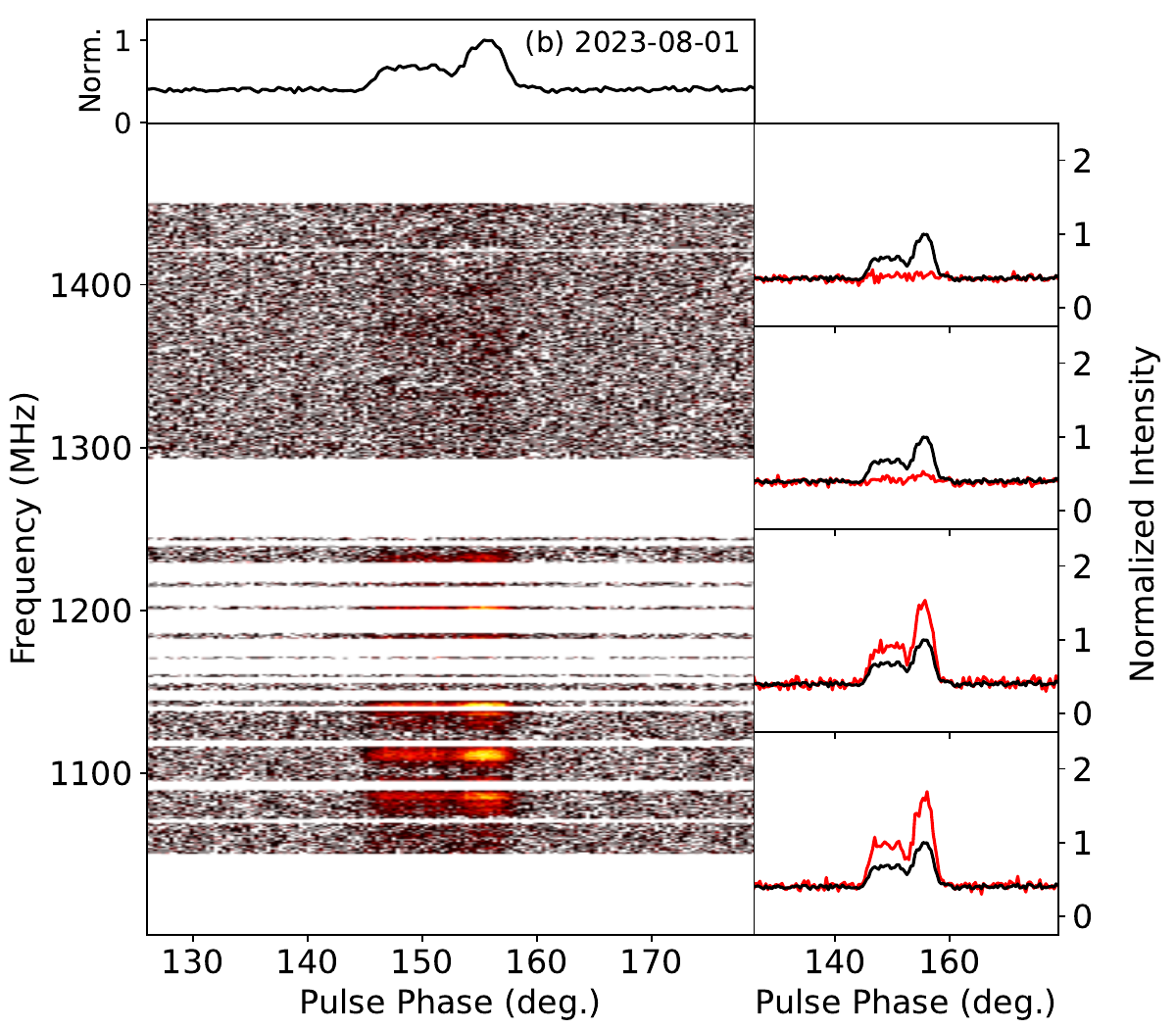}
    \caption{Frequency-dependent pulse profile variations observed at a center frequency of 1.25~GHz on (a) 2023 July 7 (left) and (b) 2023 August 1 (right). In each panel, the left subplots show the frequency–phase dynamic spectra, while the right subplots display the corresponding pulse profiles. The black curves represent the total integrated profiles across the full frequency band, while the red curves show the integrated profiles within each subband. RFI-contaminated channels have been removed and are shown in white.}

\label{freq_variations}
\end{figure*}

Following the approach of \citet{Caleb22} and \citet{Dang24}, each pulse period was divided into 8,192 phase bins, resulting in a time resolution of 49.152~$\mu$s. This high-resolution sampling allowed us to extract microstructure timescales from the time-series data of individual pulses. To characterize the temporal properties of the microstructure, we applied the autocorrelation function (ACF), a well-established technique for identifying periodic and quasi-periodic features in pulsar emission \citep{Cordes1990, Mitra2015}. In the absence of microstructure, the ACF exhibits a smooth, bell-shaped curve. However, the presence of microstructure introduces weak modulations, and when quasi-periodic components exist, the ACF displays regularly spaced secondary maxima. The time lag of the first ACF peak represents the characteristic microstructure separation ($P_\mu$), while the position of the first local minimum corresponds to the typical micro-pulse width ($\tau_\mu$).

Figure~\ref{Quasi-periodic} presents example pulses exhibiting quasi-periodic microstructures. Pulse number 1666 displays more than two quasi-periodic components, suggesting that some pulses can contain multiple quasi-periods within a single rotation. In contrast, pulse number 922 shows two distinct quasi-periodic features, while pulse number 1271 exhibits only a single quasi-periodic structure. Our analysis also suggests that quasi-periodic structures occur randomly within the sequence of regular pulses.
Across the analyzed pulses, the microstructure periodicity ranges from $P_\mu = 0.62$\,ms to $12.57$\,ms, while the width spans $\tau_\mu = 0.41$\,ms to $11.15$\,ms. On average, we find a mean microstructure period of $\langle P_\mu \rangle = 4.57$\,ms and a mean width of $\langle \tau_\mu \rangle = 4.30$\,ms. These values reflect the typical modulation characteristics within the on-pulse region.
Based on our ACF-based analysis, approximately 64\% of the regular pulses exhibit detectable microstructure features. Among these, 11.54\% demonstrate clear quasi-periodic behavior, consistent with a recent finding in the long-period pulsar PSR~J1900$-$0134, which shows 10.51\% fraction reported in the long-period pulsar PSR~J1900$-$0134 \citep{HMT2025}, where the typical microstructure timescale is approximately 2.05\,ms.

\begin{figure}
	\centering
	\includegraphics[width=\columnwidth, angle=0]{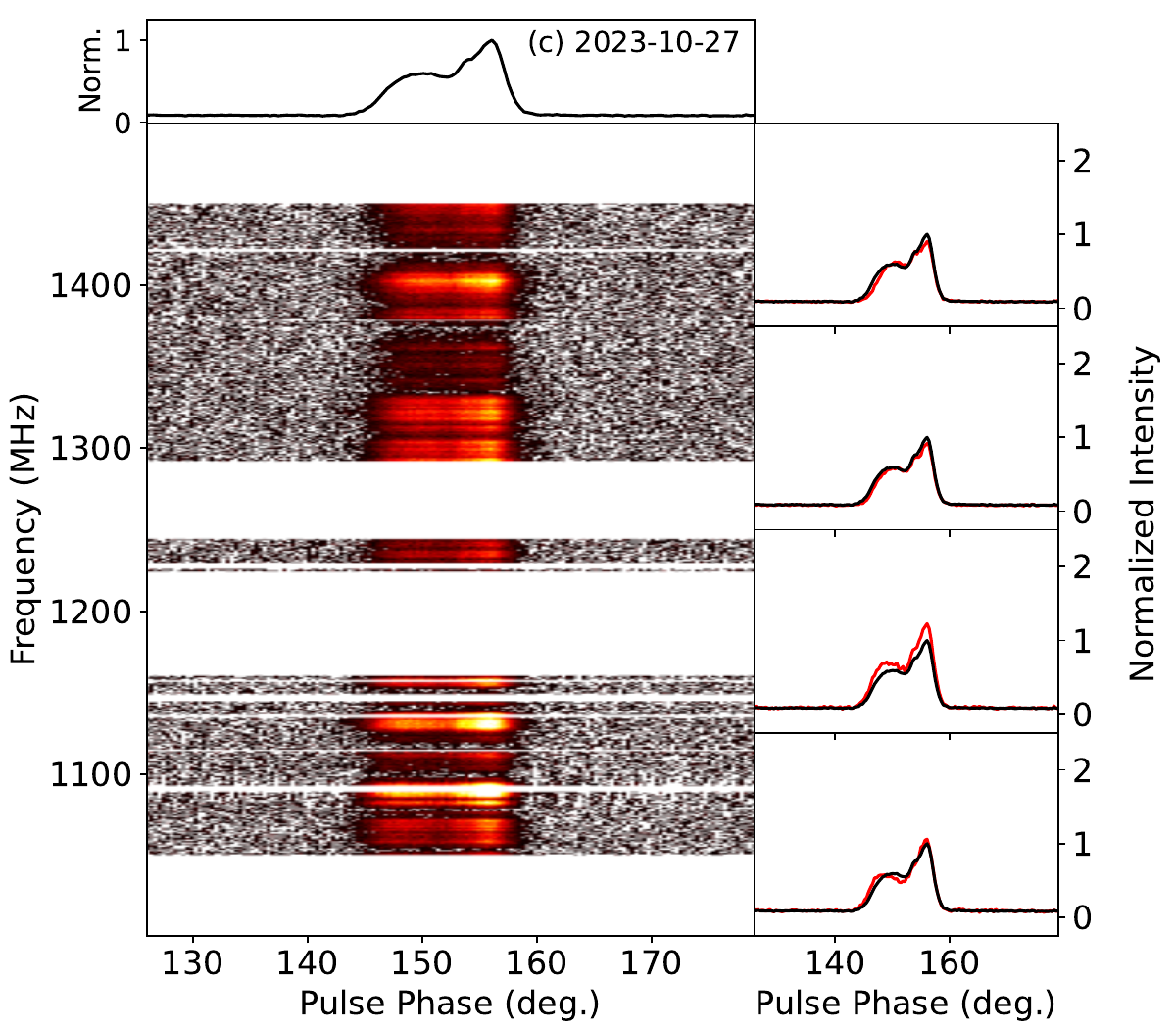}
	\caption{Similar to Figure~\ref{freq_variations}, but for the observations taken on 2023 October 27.}
\label{freq_oct27}
\end{figure}

\section{Frequency-Dependent Pulse Profile Variations} 
\label{sec:freq-dependence}

PSR~J2129+4119 is a nearby pulsar with a relatively low DM of 31\,pc\,cm$^{-3}$. According to the NE2001 model \citep{CL02}, its DM distance is estimated to be 2.3\,kpc, while the YMW16 model \citep{ymw17} suggests a slightly closer distance of 1.9\,kpc \citep{Cruces2021}. As radio waves from pulsars traverse the interstellar medium (ISM), their intensity can exhibit rapid fluctuations in both time and frequency due to diffractive interstellar scintillation (DISS; \citealt{Wang2001}).

\begin{figure}
	\centering
	\includegraphics[width=\columnwidth, angle=0]{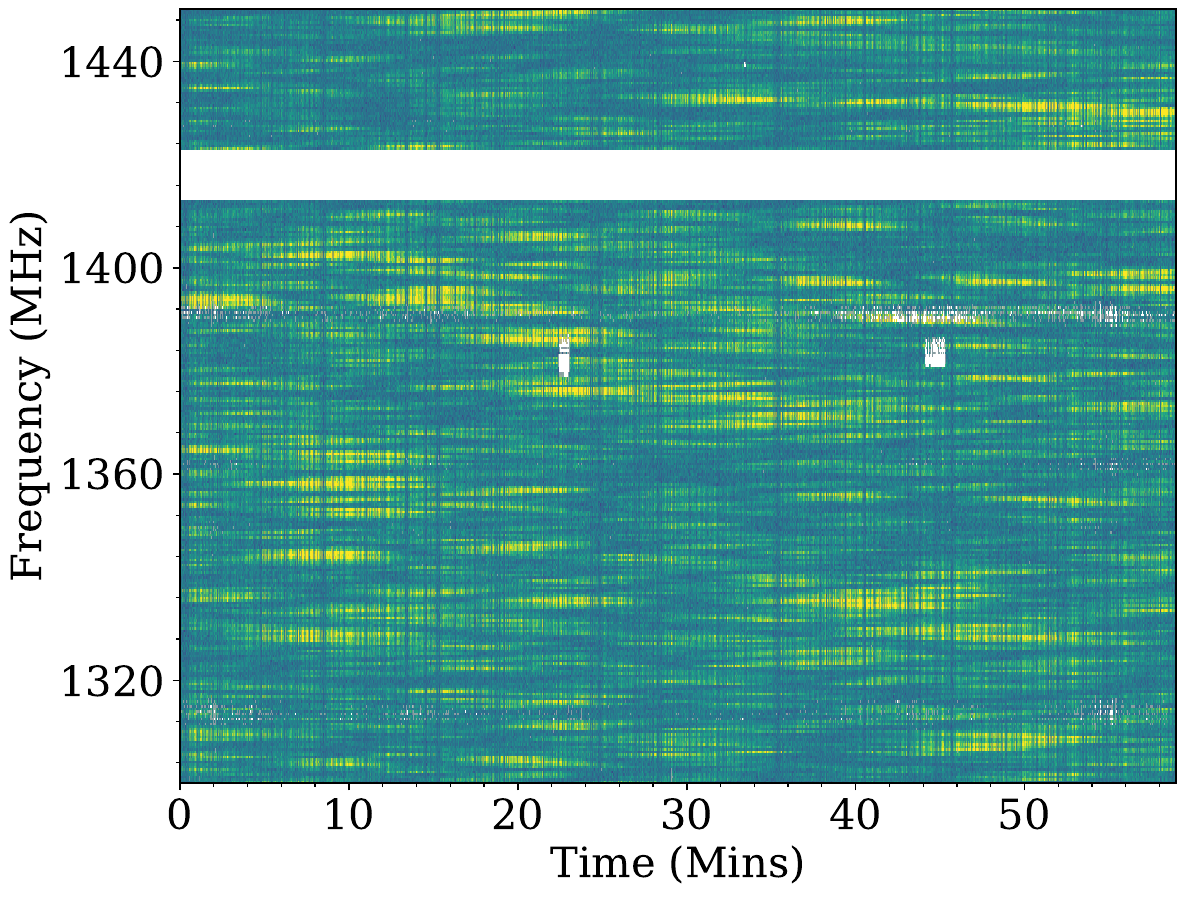}
	\caption{Frequency-time dynamic spectrum of PSR~J2129+4119 observed on 2024 September 29, with a frequency resolution of 0.488~MHz and a time resolution of 49.25~$\mu$s. The RFI-contaminated channels have been masked.}
	\label{dynamic}
\end{figure}

Figure~\ref{freq_variations} illustrates the frequency-dependent pulse profile variations of PSR~J2129+4119, based on observations conducted at a central frequency of 1.25~GHz over a bandwidth spanning 1.0–1.5~GHz. 
The two observing epochs, (a) 2023~July~7 and (b) 2023~August~1, are presented in the left and right panels, respectively. 
To examine the evolution of pulse profiles across frequency, we divided the observing band into four equal subbands within the 1.0–1.5~GHz range. 
In each subfigure, the left panels display the frequency–phase dynamic spectra, showing how the pulse intensity varies with both frequency and pulse phase, while the right panels present the corresponding integrated pulse profiles for each subband (red curves). The black curves represent the total intensity profiles integrated across the entire frequency band. The RFI-contaminated channels have been removed and are displayed in white.

The dynamic spectra reveal noticeable differences in emission characteristics as a function of frequency. In particular, some components of the profile show systematic intensity variations, while others appear to shift or broaden with increasing frequency. These changes suggest that different emission regions within the pulsar magnetosphere may dominate at different frequencies \citep{Cordes1978, Mitra2011}. Across multiple epochs, the overall pulse morphology remains stable, but small differences are visible in the relative amplitudes and widths of the leading and trailing components (Figures~\ref{freq_variations} and \ref{freq_oct27}). 
Such weak, frequency-dependent evolution is more plausibly attributed to variable propagation or scintillation effects within the magnetosphere rather than to discrete mode changing. 
In particular, the observations from 2023~July~7, August~1, and October~27 demonstrate that the pulse intensity becomes slightly more pronounced at lower frequencies, consistent with frequency-dependent absorption or birefringence in the emission region.

\begin{figure}
	\centering
	\includegraphics[width=\columnwidth, angle=0]{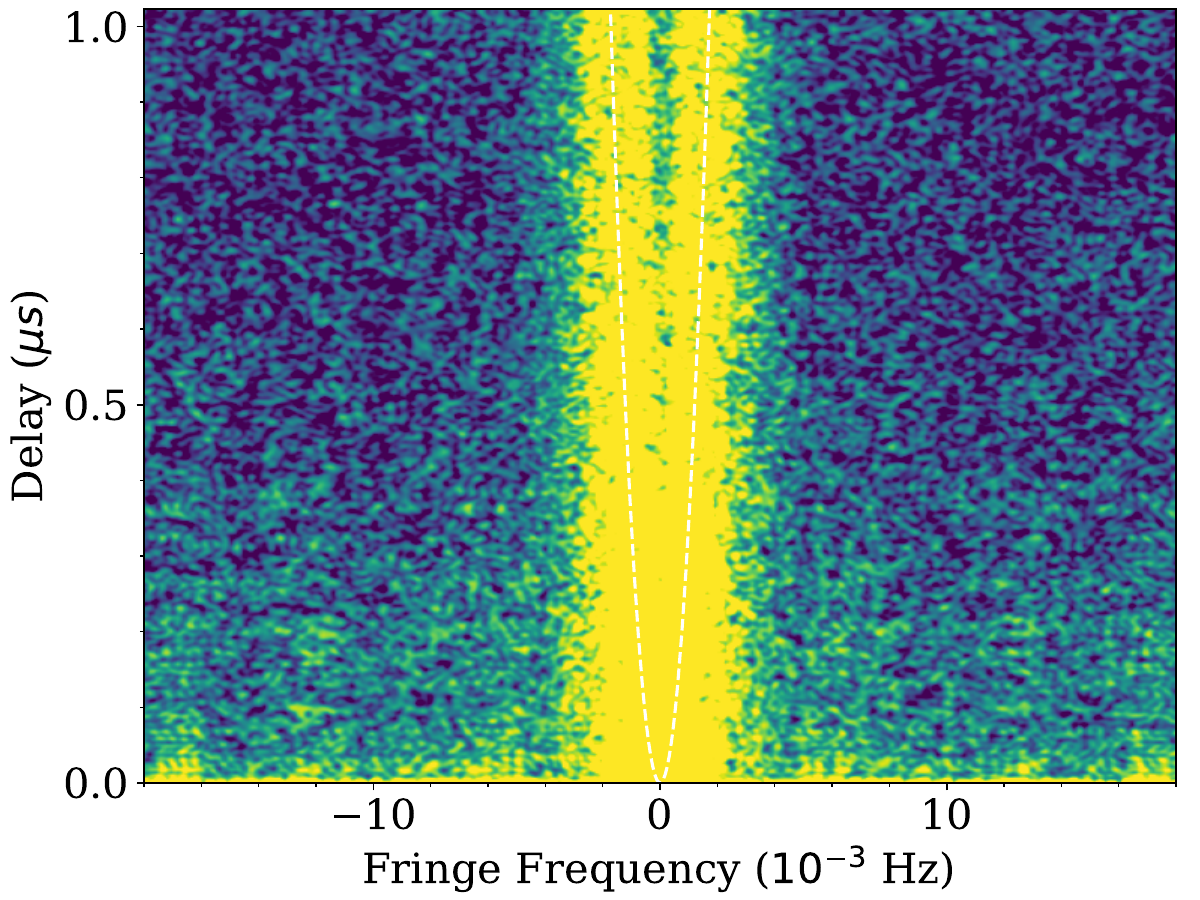}
\caption{Secondary spectrum of PSR~J2129+4119 from the 2024 September 29 observation, showing power as a function of fringe frequency and delay. A clear parabolic scintillation arc is visible, with curvature \(\eta = 0.342 \pm 0.072~\mu\mathrm{s}/\mathrm{mHz}^2\), overlaid with a dashed fit.}
	\label{sec_spec}
\end{figure}

Scintillation effects may contribute to the observed pulse profile variability of PSR~J2129+4119. As shown in Figure~\ref{dynamic}, the dynamic spectrum recorded on 2024 September 29 exhibits narrow-band intensity modulations consistent with strong diffractive interstellar scintillation. These features arise from the motion of the line of sight through a quasi-stationary interference pattern in the ISM, imprinting frequency-dependent structures onto the pulse profiles, particularly over wide observing bandwidths.

To quantify the scintillation properties, we followed the method of \citet[\texttt{scintools}; ][]{Reardon2019} and \citet{Wu2022}. We computed the two-dimensional auto-covariance function (2D ACF) and its two-dimensional Fourier transform of the secondary spectrum \citep{Stinebring2001}. The secondary spectrum provides the power distribution as a function of fringe frequency and time delay, offering insight into the scattering geometry of the ISM \citep{Walker2004, Cordes2006}. From the 2D ACF, we extracted characteristic scales using methods described in \citet{Coles2005} and \citet{Rickett2014}, yielding a scintillation bandwidth of \(\Delta\nu_\mathrm{d} = 0.89 \pm 0.02\)~MHz and a scintillation timescale of \(\Delta t_\mathrm{d} = 3.72 \pm 0.06\)~minutes, with coefficients of determination of 100.00\% and 98.79\%, respectively.

  \begin{figure}
	\centering
	\includegraphics[width=\columnwidth, angle=0]{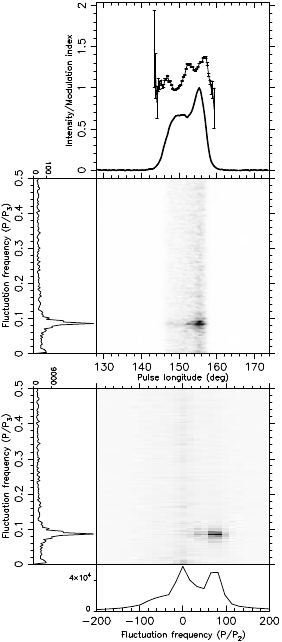}
	\caption{Fluctuation analysis of PSR~J2129+4119. Top: Integrated pulse profile with the modulation index overlaid as a solid line with error bars. Middle: Longitude-resolved fluctuation spectrum (LRFS), with the right-hand panel showing the horizontally integrated power. Bottom: Two-dimensional fluctuation spectrum (2DFS), with side panels displaying the vertically and horizontally integrated powers. A 256-point Fourier transform was used and averaged over blocks spanning the entire pulse sequence.}
	\label{2dfs_F}
\end{figure}

In the secondary spectrum shown in Figure~\ref{sec_spec}, we identify a well-defined scintillation arc, fitted using the \texttt{parabfit} routine based on the Hough transform method \citep{Bhat2016}. The best-fit arc curvature is \(\eta = 0.34 \pm 0.07~\mu\mathrm{s}/\mathrm{MHz}^2\), indicating interference between angularly scattered rays. For a thin screen, the curvature \(\eta\) is given by \citep{Cordes2006}:
\begin{equation}
\eta = 
\frac{D_{\mathrm{p}}\, s(1 - s)}{2\nu^{2}}
\frac{c}{(V_{\mathrm{eff}} \cos \psi)^{2}},
\label{eq:eta_pulsar}
\end{equation}
where \(c\) is the speed of light, \(\nu\) the observing frequency, \(D_{\mathrm{p}}\) the pulsar distance, \(s\) the fractional screen distance, \(\psi\) the angle between the scattering structure and the effective velocity \(V_{\mathrm{eff}}\).
Assuming \(V_{\mathrm{eff}} = 300~\mathrm{km\,s^{-1}}\), \(D_{\mathrm{p}} = 2.3~\mathrm{kpc}\), \(\nu = 400~\mathrm{MHz}\), and \(\cos \psi = 1\), we estimate a fractional screen distance of \(s = 0.04 \pm 0.01\), corresponding to a physical screen distance of \(d_{\mathrm{scr}} \approx 0.1~\mathrm{kpc}\) from the observer. The near-observer solution places the scattering region within or near the boundary of the Local Bubble — the low-density cavity surrounding the Sun that extends to approximately 100–200~pc and whose interface with denser interstellar material is known to host enhanced electron-density irregularities \citep{Lallement2003,Alves2020}. Such a location is consistent with a thin, localized structure in the nearby ionized interstellar medium. 
However, a more precise determination of the screen location would require knowledge of the velocity components of the scattering medium in right ascension and declination, the orientation angle between the scattering structure and the effective velocity vector, the pulsar's proper motion, and a precisely measured distance. Although timing observations are available, the current data span is insufficient for a reliable proper-motion measurement and for long-term scintillation studies. Future timing and interferometric observations will therefore be essential to refine the velocity model and to localize the scattering screen more accurately.

\section{Subpulse drifting}
\label{sect:subpulse_drifting}

As seen in the pulse stack (Figure~\ref{stack1}), PSR~J2129+4119 exhibits regular subpulse drifting in the forward (increasing longitude) direction. To investigate this behavior quantitatively, we performed a fluctuation spectral analysis using \textsc{PSRSALSA}\footnote{https://github.com/weltevrede/psrsalsa}\citep{Weltevrede16}, the results are presented in Figure~\ref{2dfs_F}.

\subsection{Subpulse fluctuation analysis}\label{subpulse-drifting}

The top panel of Figure~\ref{2dfs_F} displays the integrated pulse profile along with the longitude-resolved modulation index, which quantifies the relative intensity variations across pulse phase. The modulation index \( m_i = \sigma_i / \mu_i \), where \( \sigma_i \) is the standard deviation and \( \mu_i \) is the mean intensity at longitude bin \( i \) \citep{Weltevrede06},  peaks in the trailing component of the profile, indicating strong subpulse intensity fluctuations in that region. In contrast, the bridge region between the leading and trailing components shows a clear dip in modulation index, suggesting relatively stable emission there. The leading component displays lower modulation overall but is associated with noticeably larger error bars, consistent with increased uncertainty also observed in the polarization angle measurements (see Figure~\ref{RVM_fit}). The error bars shown in the modulation index curve were estimated using a bootstrapping technique, where Gaussian noise equivalent to the RMS of the off-pulse region was added in each iteration.

To characterize the periodicity and drift behavior, we computed the longitude-resolved fluctuation spectrum (LRFS; \citet{Edwards03}) as shown in the middle panel of Figure~\ref{2dfs_F}. Null pulses were excluded from the sequence to isolate regular phases, and the data were divided into blocks of 256 pulses. A 256-point discrete Fourier transform was applied to each phase bin, and the resulting spectra were averaged to produce the final LRFS. This analysis reveals a clear modulation feature, indicating a well-defined subpulse modulation period (\( P_3 \)).

To assess subpulse movements across pulse longitude, we also computed the two-dimensional fluctuation spectrum (2DFS; \citet{Edwards02}), as shown in the bottom panel of Figure~\ref{2dfs_F}. The presence of a drift feature offset from the vertical axis confirms the existence of longitudinal subpulse drift, with a measurable horizontal separation corresponding to a drift phase step (\( P_2 \)). In both LRFS and 2DFS, the vertical axis represents the fluctuation frequency \( P/P_3 \), while in 2DFS, the horizontal axis corresponds to \( P/P_2 \), where \( P \) is the pulsar rotation period. Visual inspection of the 2DFS reveals a modulation peak at approximately \( 0.085 \, P/P_3 \), corresponding to a subpulse modulation period of roughly \( P_3 \approx 11.76\,P \). A rough estimate of the subpulse separation \( P_2 \) can also be made from the horizontal offset of the drift feature, which appears near \( P/P_2 \approx 70 \). This corresponds to a longitude separation of \( P_2 \approx 5^{\circ}.14 \). To obtain a more precise measurement, we applied the \texttt{pspecDetect} routine from the \textsc{PSRSALSA} package, which yielded \( P_3 = 11.74 \pm 0.03\,P \) and \( P_2 = 7^{\circ}.00 \pm 0^{\circ}.31 \), respectively. These results confirm the presence of regular subpulse drifting in PSR~J2129+4119 with well-defined periodicities in both phase and longitude.

\begin{figure}
	\centering
	\includegraphics[width=\columnwidth, angle=0]{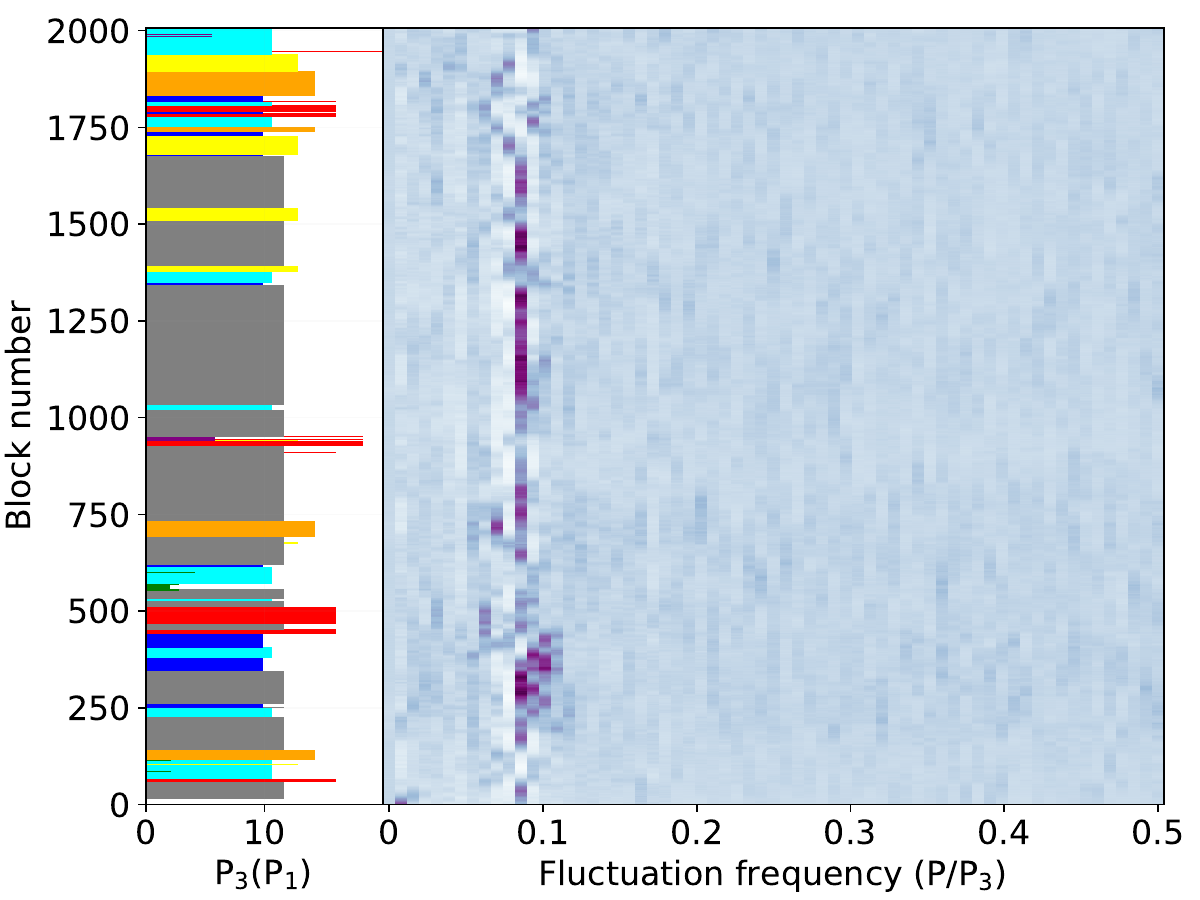}
	\caption{The Sliding two-dimensional fluctuation spectrum (S2DFS) for PSR~J2129+4119 observed on 2024 September 29.}
	\label{S2DFS}
\end{figure}

\begin{table*}
	\setlength{\tabcolsep}{26 pt}
	\centering
	\caption{Extracted fluctuation parameters ($P_3$ and $P_2$) from different segments of the observation.}
	\label{tab:fluctuation_summary}
	\begin{tabular}{ccccc}
		\hline	\hline
		\#Panel & 	\#Block & $P_3$ (P) & $P_2$ ($^\circ$) & Drift mode \\
		\hline
		Left   & 550--800      & $11.74 \pm 0.03$ & $ 6.99 \pm 0.31$  & Single drift mode \\
		Middle  & 300--500   & $14.24 \pm 0.07$ & $6.13 \pm 0.35$  & Double drift mode \\
		&& $11.66 \pm 0.06$ & $6.57 \pm 0.50$  & \\
		Right & 1700--2000   & $13.51 \pm 0.03 $  & $5.27 \pm 0.65$  & Beat-like features \\
		\hline
	\end{tabular}
	
	\vspace{2mm}
	\noindent
	\begin{minipage}{\textwidth}
		\small \textbf{Note.} \#Panel refers to the location of the plots for each 2DFS fluctuation analysis in Figure~\ref{fluctuation_each}, while \#Block represents the block numbers used in the S2DFS analysis to estimate the evolution of $P_3$ values over time, as shown in Figure~\ref{S2DFS}.
	\end{minipage}
	\label{tab:fluctuation}
\end{table*}

\begin{figure*}
	\centering
	\includegraphics[width=0.68\columnwidth, angle=0]{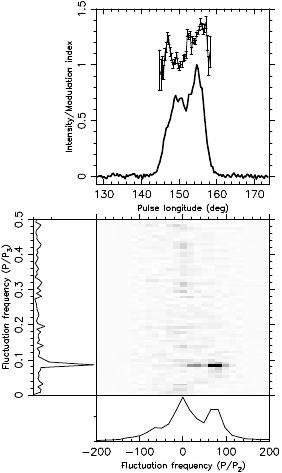}
	\includegraphics[width=0.68\columnwidth, angle=0]{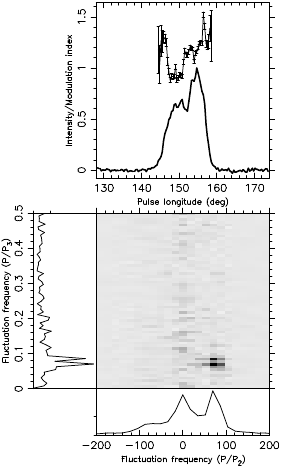}
	\includegraphics[width=0.68\columnwidth, angle=0]{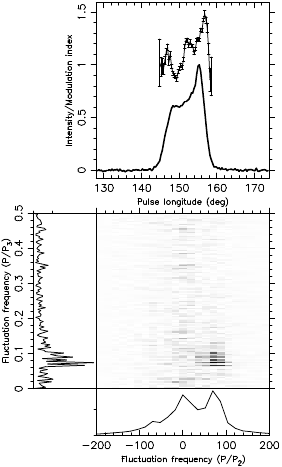}
	\caption{Similar to Figure \ref{2dfs_F}, but for each fluctuation analysis for Figure \ref{stack1}. }
	\label{fluctuation_each}
\end{figure*}

\subsection{Beat-like modulation features}

The S2DFS was computed by dividing the pulse sequence into overlapping blocks of 128 pulses, each shifted by one pulse period. For each block, we calculated the 2DFS and collapsed it along the $P_2$ axis to obtain a one-dimensional fluctuation spectrum. These spectra were then stacked to form the time-evolving S2DFS, as shown in Figure~\ref{S2DFS}. The left panel illustrates the evolution of $P_3$ values over time, where each color represents a different fluctuation frequency. 

Figure~\ref{fluctuation_each} presents the 2DFS analysis for the segmented pulse sequences shown in Figure~\ref{stack1}. The panels reveal distinct $P_3$ and $P_2$ values across different intervals, along with notable changes in modulation behavior. In the left panel, a single drift mode is observed with $P_3 = 11.74 \pm 0.03 \,P$ and $P_2 = 6^{\circ}.99 \pm 0^{\circ}.31$, indicative of regular subpulse modulation.

The middle panel shows two prominent fluctuation peaks, suggesting the coexistence of multiple drift modes, with periodicities at $P_3 = 14.24 \pm 0.07\,P$ and $11.66 \pm 0.06\,P$, and corresponding $P_2$ values of $6^{\circ}.13 \pm 0^{\circ}.35$ and $6^{\circ}.57 \pm 0^{\circ}.50$, respectively. These features are particularly noticeable in the block range of approximately 300–620 in Figure~\ref{S2DFS}.

The right panel exhibits more complex behavior, where multiple fluctuation tracks appear and shift from higher to lower frequencies (i.e., shorter $P_3$), showing a pattern such as 0.085, 0.080, 0.075, 0.070, and back again. This could signify beat-like modulation features, which are evident in the block range 1700–2000 in Figure~\ref{S2DFS}. The corresponding 2DFS shows multiple components, including a strong feature at $P_3 = 13.51 \pm 0.03\,P$ with $P_2 = 5^{\circ}.27 \pm 0^{\circ}.65$. A summary of these features is provided in Table~\ref{tab:fluctuation}.

\section{Discussion} \label{sec:discus}

\subsection{Emission height and geometry}

The polarization behavior of PSR~J2129+4119 offers valuable insights into its emission geometry. The smooth S-shaped position angle (PPA) swing (Figure~\ref{RVM_fit}) is well-described by the Rotating Vector Model \citep[RVM;][]{RadhakrishnanCooke1969}, consistent with a dipolar magnetic field configuration. RVM fitting applied to both the full dataset and the regular emission mode yields consistent geometrical parameters (Table~\ref{tab:rvm_fit}), with overlapping PPA curves indicating a stable magnetospheric geometry.
Despite the good fits, a significant degeneracy persists between $\alpha$ and $\beta$ (Figure~\ref{alpha_beta_contour}), primarily due to the small $|\beta|$ and the limited phase span of PPA points \citep{Everett01, Rookyard2015}. The small $|\beta|$ suggests that the observer’s line of sight passes close to the magnetic axis, which is also supported by the near-symmetric profile shape \citep{Helfand1975}.

The steepest gradient (SG) in the PPA curve occurs at $\phi_0 \approx 151^\circ$, slightly preceding the centroid of the intensity profile. This offset is attributed to aberration and retardation (A/R) effects \citep{Blaskiewicz1991, Mitra2015}, implying that the radio emission originates from a significant height above the neutron star surface. The A/R-induced phase delay between the SG point and the profile center is expressed as:
\begin{equation}
	\Delta\phi = \phi_0 - \phi_c,
\end{equation}
where $\phi_0$ is the RVM-fitted SG phase, and $\phi_c$ is the intensity centroid. The corresponding emission height $h_{\mathrm{em}}$ is then estimated by \citep{Blaskiewicz1991}:
\begin{equation}
	h_{\mathrm{em}} = \frac{P c}{8\pi \Delta\phi},
	\label{eq:hem_inverse}
\end{equation}
where $P$ is the pulsar period and $c$ is the speed of light.
Assuming $\Delta\phi$ is the only parameter with significant uncertainty, the error propagation yields the uncertainty in $h_{\mathrm{em}}$ as:
\begin{equation}
	\delta h_{\mathrm{em}} = \frac{P c}{8\pi} \cdot \frac{\delta(\Delta\phi)}{(\Delta\phi)^2},
	\label{eq:hem_uncertainty}
\end{equation}
where $\delta(\Delta\phi)$ is the uncertainty in the measured phase offset.
For PSR~J2129+4119, adopting $P = 1.69$~s and $\Delta\phi = 0^\circ.51$, we obtain an emission height of $h_{\mathrm{em}} = 178.32 \pm 14.05$~km.
Assuming the emission originates from the last open dipolar field lines, this height corresponds to a radial distance of approximately $0.23\%$ of the light cylinder radius, measured from the center of the neutron star. This result indicates that, even in long-period pulsars with moderate inclination angles and emission heights, such as PSR~J2129+4119, the effects of aberration and retardation can be significant.
A similar behavior has recently been reported in another long-period pulsar discovered by FAST, PSR~J0344$-$0901 \citep{HMT2024}, further supporting the importance of emission height and A/R effects on the profile asymmetry and polarization characteristics of such pulsars.

The emission height can also be related to the half-opening angle $\Gamma$ of the emission cone using \citep{Rankin90}:
\begin{equation}
	\Gamma = \sqrt{\frac{9\pi h_{\mathrm{em}}}{2 P c}}.
	\label{eq:rho_height}
\end{equation}
For the estimated $h_{\mathrm{em}}$, we find $\Gamma \approx 4^\circ.04 \pm 0^\circ.16$.
The beam opening angle $\Gamma$ is geometrically related to the observed pulse width $W_{\mathrm{open}}$ through the relation \citep{Gil1984}:
\begin{equation}
	\cos\Gamma = \cos\alpha \cos\zeta + \sin\alpha \sin\zeta \cos\left( \frac{W_{\mathrm{open}}}{2} \right),
	\label{eq:rho_width}
\end{equation}
where $W_{\mathrm{open}} = 13^\circ.71$ is the pulse width at the 10\% intensity level. 
Equations~\ref{eq:rho_height} and \ref{eq:rho_width} show that measurements of $\Gamma$ and $W_{\mathrm{open}}$, together with RVM constraints, can be used to refine further estimates of $\alpha$ and $\beta$. The gray-shaded region in Figure~\ref{alpha_beta_contour} marks $(\alpha, \beta)$ combinations excluded by these geometric constraints.

Notably, the polarimetric properties differ significantly between emission modes. The regular mode exhibits stable and moderately high linear polarization ($\langle L \rangle / \langle I \rangle \approx 51\%$) with weak circular components, indicating well-organized polarization behavior. In contrast, the weak mode shows a marked reduction in linear polarization ($\sim$25\%), consistent with its lower signal-to-noise level. Although the weak mode displays an apparently large $\langle |V| \rangle / \langle L \rangle$ ratio, this likely reflects residual noise effects rather than intrinsic circular-polarization dominance.

\begin{figure}
	\centering
	\includegraphics[width=\columnwidth, angle=0]{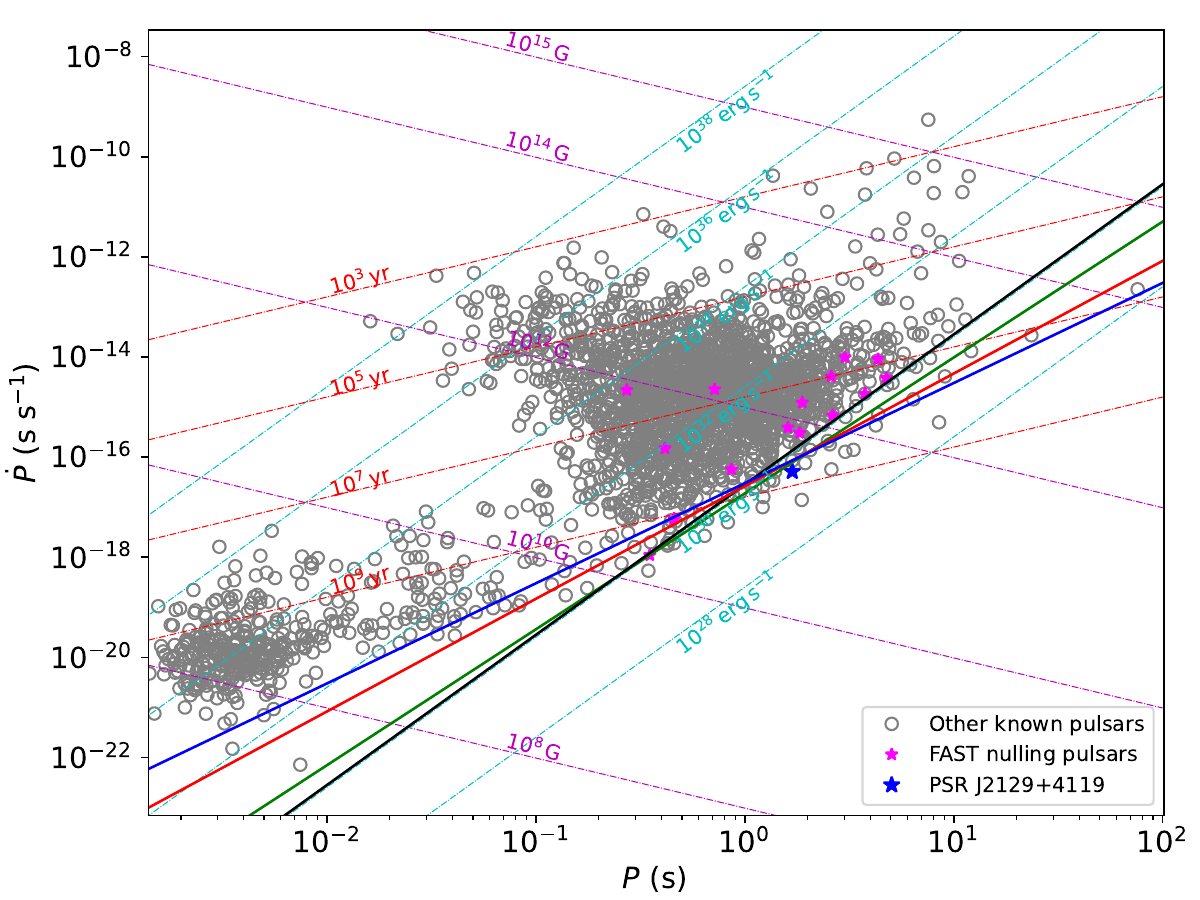}
    \caption{$P$–$\dot{P}$ diagram for 3474 pulsars. Gray circles mark known pulsars, magenta stars indicate FAST nulling pulsars, and blue star–highlighted circles show our candidates for PSR~J2129+4119. Diagonal lines denote constant magnetic fields (magenta), spin-down luminosities (cyan), and characteristic ages (red). The lower-right corner includes the traditional death line (black; \citealt{Ruderman1975}), the death valley (red; \citealt{Chen1993}), and curvature-radiation limits from the vacuum-gap (green) and space-charge-limited flow (blue) models \citep{Zhang2000}. Data are from the ATNF pulsar catalogue \citep{Manchester05}.}
	\label{ppdot}
\end{figure} 
 
\subsection{PSR~J2129+4119 and the pulsar population}\label{pulsar-populations}

The radio pulsar ``death line'' represents a theoretical boundary in the $P$–$\dot{P}$ diagram, below which pair production is insufficient to sustain coherent radio emission \citep{Ruderman1975, Chen1993, Zhang2000}. The traditional death line, derived from the vacuum gap model, corresponds to a spin-down luminosity of $\dot{E} \approx 1.53 \times 10^{30}$\,erg\,s$^{-1}$ \citep{Wu2023}. Pulsars with $\dot{E}$ below this threshold, commonly referred to as low-$\dot{E}$ pulsars, are generally expected to be radio-quiet or exhibit sporadic, weak emission \citep{Szary2014}.

PSR~J2129+4119, with $\dot{E} \approx 0.65 \times 10^{30}$\,erg\,s$^{-1}$ and a characteristic age of 342.8\,Myr, lies not just the traditional death line but also the extended death lines proposed by \citet{Zhang2000}. Its position in the $P$–$\dot{P}$ diagram (Figure~\ref{ppdot}) places it among a small subset of known pulsars with such low energy loss rates. Despite this, our FAST observations reveal sustained and multi-modal emission behavior, including three distinct modes: nulls, weak pulses, and regular pulses. The pulsar also occasionally emits bright pulses, quasi-periodic microstructure, and displays clear subpulse modulation features. 

Interestingly, many pulsars near the death line exhibit significantly higher nulling fractions. For example, the FAST-discovered nulling pulsars PSR~J2323+1214 and PSR~J1945+1211, both located close to the death line, have measured NFs of 49\% and 53\%, respectively \citep{HMT2025}. This trend is also observed in other low-$\dot{E}$ pulsars discovered by FAST, such as PSR~J1611$-$0114 (NF = 40\%, $\dot{E} = 9.2 \times 10^{30}$,erg,s$^{-1}$; \citealt{Liu2024}) and PSR~J0211+4235 (NF = 49\%, $\dot{E} = 1 \times 10^{30}$,erg,s$^{-1}$; \citealt{Guo2023}). The high nulling fractions in these sources are consistent with reduced pair production efficiency near the polar caps, as expected when pulsars approach the death line. In contrast, the relatively low NF and persistent emission of PSR~J2129+4119 challenge conventional expectations for pulsars beyond the death line. This suggests that coherent radio emission can persist even at extremely low $\dot{E}$, potentially due to localized magnetic field enhancements \citep{Gil2003}, favorable emission geometry \citep{Basu16}, or residual pair production \citep{Zhang2000}. 

\citet{Chen1993} proposed that twisted magnetic field lines at the polar cap could sustain pair production in extreme cases. This is supported by observations in PSR~J0211+4235 \citep{Wu2023}, where twisted fields enable strong pair production. Similar fields are observed in other pulsars, including PSR~J0815+0939 \citep{Szary2017}. Comparisons with pulsars like PSR~J2144--3933 \citep{Young1999}, PSR~J0250+5854 \citep{Tan2018}, and PSR~J0211+4235 \citep{Wu2023}, all beyond the death line, suggest that twisted magnetic fields may enable pair production. Thus, the persistence of pair production in PSR~J2129+4119 could also be due to twisted magnetic fields.

Subpulse drifting and amplitude modulation are not uniformly distributed across the pulsar population. \citet{Basu16} showed that drifting occurs mostly in pulsars with $\dot{E} < 5 \times 10^{32}$,erg,s$^{-1}$, with $P_3$ anti-correlated with $\dot{E}$, while higher-$\dot{E}$ pulsars tend to exhibit phase-stationary amplitude modulation. A larger survey by \citet{Song2023} found drifting in 35\% and pure amplitude modulation in only 15\% of 1198 pulsars, with coexisting features being rare.
PSR~J2129+4119 shows predominantly phase-modulated subpulse drifting, with stable $P_3$ and $P_2$ values across most of the observation (e.g., $P_3 \approx 11.74,P$, $P_2 \approx 6^\circ.99$). However, in certain intervals, particularly between blocks 1700 and 2000, the fluctuation spectra reveal more complex modulation features. These include multiple coexisting drift components (e.g., $P_3 = 13.51 \pm 0.03,P$, $P_2 = 5^\circ.27 \pm 0^\circ.65$) and quasi-periodic shifts in modulation frequency, forming symmetric patterns in the S2DFS. These structures are consistent with beat-like modulation effects, likely arising from interference between closely spaced drift modes. Unlike pulsars such as PSR~J1514$-$4834 \citep{Li2024, Hsu2025}, which exhibit fast amplitude modulation (flickering) alongside slow drift, PSR~J2129+4119 remains drift-dominated, with beat-like behavior appearing intermittently rather than as a defining characteristic.

\subsection{Implications of the emission mechanisms}

The radio emission from pulsars is widely believed to originate from regions near the magnetic poles, where strong electric fields in vacuum or partially screened gaps accelerate charged particles to relativistic speeds \citep{Melrose14, Timokhin2019, Philippov2020}. These particles emit coherent radiation, predominantly via curvature radiation, as they follow curved magnetic field lines \citep{Ruderman1975}. The sustainability and variability of such emissions are closely linked to the efficiency of plasma generation in the pulsar magnetosphere  \citep{Melrose14}.

Our FAST observations of PSR~J2129+4119 reveal a range of emission states, nulls, weak pulses, regular pulses, occasional bright bursts, and quasi-periodic microstructure, all indicative of a dynamic and unstable magnetospheric environment. The emission is not only variable from pulse to pulse but also shows systematic changes across different states. Notably, the integrated pulse profiles before and after nulls exhibit asymmetries: regular emission (RFAP) is significantly stronger after nulls, especially in the trailing component, while weak pulses show greater instability in profile shape and strength across null boundaries. These variations imply a temporary disruption and reactivation of plasma flow, likely linked to changes in pair production efficiency or magnetospheric current structure.

Several mechanisms have been proposed to explain such nulling and emission variability in pulsars. These include the failure of pair production in the polar cap region \citep{Jones1981}, temporary loss of coherence among relativistic particles \citep{Filippenko1982}, switching between curvature radiation and inverse Compton scattering \citep{Zhang1997}, and reconfiguration of the global magnetosphere \citep{Timokhin10}. Geometric effects, such as the line of sight missing the emission region due to field line rearrangement, have also been considered \citep{Herfindal2007}, as have stellar oscillations \citep{Rosen2011} and magnetic field instabilities \citep{Geppert2021}.
In our case, the rapid recovery of regular emission after nulls, accompanied by a more intense and structured trailing component, points toward a gradual magnetospheric reactivation rather than a purely geometric origin. The fact that the null durations observed in PSR~J2129+4119 are short, typically lasting fewer than four pulse periods, further supports this interpretation. Furthermore, the high degree of polarization observed in both regular and weak pulses, combined with consistent PPA swings as described in the RVM model, suggests that the emission region remains well-aligned and stable during active phases.

Additionally, interstellar scintillation may contribute to some of the observed pulse-to-pulse variability. The secondary spectrum of PSR~J2129+4119 shows a well-defined scintillation arc with curvature $\eta = 0.342 \pm 0.072~\mu\mathrm{s}/\mathrm{MHz}^2$, indicative of interference between angularly scattered wavefronts. This points to a thin, localized scattering screen along the line of sight and underscores the role of propagation effects in modulating the received signal \citep{Stinebring2001, Bhat2016, Wu2022}.
Together, these findings emphasize that both intrinsic (e.g., magnetospheric activity, plasma instabilities) and extrinsic (e.g., scintillation) factors influence the emission variability in PSR~J2129+4119. The pulsar’s diverse observational behavior, including beat-like modulation, emission asymmetries near nulls, and quasi-periodic microstructure, suggests that its magnetosphere operates near the threshold for coherent emission, providing valuable constraints on the physical conditions that enable or suppress pulsar radio emission.

\section{Summary}\label{sec:Summary}

We have conducted the first detailed single-pulse study of PSR~J2129+4119, based on a one-hour FAST observation on 2024 September 29 at a central frequency of 1.25\,GHz. The pulsar has a rotation period of $1.69$\,s and a low dispersion measure of 31\,cm$^{-3}$\,pc, making its emission strongly susceptible to scintillation effects. Our main results are as follows:

\begin{enumerate}
	\item PSR~J2129+4119, with a spin-down luminosity $\dot{E} \approx 0.65 \times 10^{30}$\,erg\,s$^{-1}$, lies well below the traditional death line, yet exhibits sustained and multi-modal emission behavior.
	\item The pulsar exhibits different emissions that can be classified into nulls, weak pulses, and regular pulses from pulse-energy distributions. The measured nulling fraction is $8.13\% \pm 0.51\%$.
	\item Fluctuation spectral analyses (2DFS and S2DFS) reveal both stable phase-modulated drifting and beat-like modulation, the latter most prominent between pulse numbers 1700–2000.
	\item Both regular and weak pulses show high linear polarization. RVM fitting yields a small impact angle ($\beta \approx -3^\circ$), consistent with a near-tangential line of sight.
	\item Post-null regular pulses display enhanced trailing components relative to pre-null pulses, suggesting gradual magnetospheric reactivation rather than a purely geometric origin.
	\item Microstructure is present in $\sim$ 64\% of regular pulses, with mean periodicity $\langle P_\mu \rangle = 4.57$\,ms and mean width $\langle \tau_\mu \rangle = 4.30$\,ms.
	\item Dynamic and secondary spectra reveal strong diffractive interstellar scintillation. The measured arc curvature, $\eta = 0.342 \pm 0.072~\mu$s/MHz$^2$, implies a thin, localized scattering screen.
	\item The coexistence of drifting, beat-like modulation, and stable polarization geometry indicates that coherent emission can persist in low-$\dot{E}$ pulsars near the pair-production threshold.
\end{enumerate}

\section*{Acknowledgments}

This work is supported by the National Natural Science Foundation of China (NSFC)(Grant Nos. 12588202, U2031117, U1838109, 11873080, 12041301) and by the Alliance of International Science Organizations, Grant No. ANSO-VF-2024-01. We are grateful to the anonymous referee for valuable comments that have improved the presentation of this paper. H.M.T. acknowledges Arba Minch University. We thank Dr. Ralph Eatough for useful discussions. D.L. is a new cornerstone investigator and is supported by NSFC (12588202) and the 2020 project of Xinjiang Uygur Autonomous Region of China for flexibly fetching in upscale talents. P.W. acknowledges support from the NSFC Programs (12588202, 12041303), the CAS Youth Interdisciplinary Team, the Youth Innovation Promotion Association CAS (id. 2021055), and the Cultivation Project for FAST Scientific Payoff and Research Achievement of CAMS-CAS. R.Y. is supported by the National Natural Science Foundation of China (NSFC) project (No. 12041303, 12041304, 12288102), the National Key Program for Science and  Technology Research and Development No. 2022YFC2205201, the National SKA  Program of China No. 2020SKA0120200, the Major Science and Technology Program of Xinjiang Uygur Autonomous Region No. 2022A03013--2, 2022A03013--4. This research is partly supported by the Operation, Maintenance, and Upgrading Fund for Astronomical Telescopes and Facility Instruments, budgeted from the Ministry of Finance of China (MOF) and administered by the CAS. We also thank all the members of the FAST telescope collaboration for their efforts in establishing the projects (project number: PT2024$_{-}$0036), which made these observations possible. 

This work made use of data from the Five-hundred-meter Aperture Spherical radio Telescope (FAST). FAST is a Chinese national mega-science facility operated by the National Astronomical Observatories, Chinese Academy of Sciences.

\software{DSPSR \citep{Straten11}, PSRCHIVE \citep{Hotan04}, TEMPO2 \citep{Hobbs06}, PSRSALSA \citep{Weltevrede16}}

\bibliography{Multi_Faceted_Emission_Properties_of_PSR_J2129_4119_Observed_with_FAST}{}
\bibliographystyle{aasjournalv7}

\end{document}